\documentclass[fleqn,12pt]{article}
\usepackage{latexsym,amssymb,amsmath}

\newcommand{\p}{\partial}

\begin{document}

\allowdisplaybreaks

\begin{center}
\Large \bf
Generalized Killing Tensors and Symmetry \\ of Klein--Gordon--Fock
Equations\footnote{The first version of the paper was published
as a preprint in Russian: Preprint N 90.23, Kyiv, Institute of Mathematics,
1990, 59~p.}
\end{center}

\begin{center}
\large \bf Anatoly G. Nikitin and  Oleksander I. Prylypko
\end{center}

\begin{center}
{\it Institute Institute of Mathematics of NAS of Ukraine,\\
 3 Tereshchenkivs'ka Str., Kyiv, 01601 Ukraine}\\
E-mail: nikitin@imath.kiev.ua, URL: http://www.imath.kiev.ua/\~{}nikitin/
\end{center}

\begin{abstract}
The paper studies non-Lie symmetry of the Klein--Gordon--Fock
equation (KGF) in $(p+q)$-dimensional Minkowsky space. Full set of
symmetry operators for the $n$-order KGF equation was explicitly
calculated for arbitrary $n<\infty$ and $p+q \leq 4$.

Definition was given for generalized Killing tensors of rank $j$
and order $s$, and for generalized conformal Killing tensors of
rank $j$ and order $s$ as a complete set of linearly independent
solutions of some overdetermined systems of PDE. These tensors
were found in explicit form for arbitrary fixed $j$ and $s$ in
Minkowsky space of dimension $p+q \leq 4$. The received results
can be used in investigation of higher symmetries of a wide class
of systems of partial differential equations.
\end{abstract}

\section*{Introduction}

Classical group theoretical analysis of differential equations
whose foundations were laid by Sophus Lie over a hundred years ago
finds increasing utilization in modern mathematical physics (see
e.g. ~\cite{Ovsyannikov,Ibragimov,Olver}). At the same time
certain limits of the classical Lie approach become obvious that
nevertheless do not allow full description of the symmetry of an
equation under study \cite{F1, F2}. In particular, does
not allow calculation of higher order symmetry operators that are
widely used for calculation of reference frames admitting solution
of equations in separated variables \cite{Shapovalov,Miller,Kalnins},
in calculation of motion constants~\cite{F3} and
in many other problems.

The present paper deals with investigation of non-Lie symmetry of
the Klein--Gordon--Fock equation in $(p+q)$-dimensional Minkowsky
space
\begin{gather}
 L\varphi \equiv\left( g^{\mu\nu}\frac{\p}{\p x_\mu}
\frac{\p}{\p x_\nu}-\varkappa^2\right)\varphi=0,
\end{gather}
where $\varkappa$ is a real parameter,
\begin{gather} g^{\mu\nu}=\left\{\begin{array}{rl} 0, &  \mu \not= \nu, \\
1,  &  \mu=\nu \leq p, \\
-1, &  p < \mu=\nu \leq p + q, \end{array} \right.
\end{gather}
$\varphi=\varphi(x_1, x_2, \ldots, x_{p+q})$ is a function of
$p+q$ variables.

A symmetry operator of equation (1) is an arbitrary operator $Q$
(linear, nonlinear, differential, integral) that transforms
solutions of this equation into solutions, that is \cite{F3}
\begin{gather}
 L(Q \varphi)=0, \quad \mbox{if} \quad L \varphi=0
\end{gather}
(see Section 1 below for more rigorous definition).

We will call a differential operator of a finite order $n$ being a
symmetry operator of equation (1) a $n$-th order symmetry
operator.

Description of the maximal (in the sense of Lie) symmetry of
equation (1) may be reduced to finding of all linearly independent
first order symmetry operators. Such operators are well-known,
they form a basis of the Lie algebra of the generalized Poincar\'e
group $P(p,q)$ (for $\varkappa \neq 0$) or for the conformal group
in $(p+q)$-dimensional space (when $\varkappa = 0$).

One of the main results of the present paper is calculation in
explicit form of a complete set of $n$-th order symmetry operators
of equation (1) for arbitrary $n < \infty$ and $p+q\leq 4$.

It is well-known that description of first order symmetry
operators is based upon calculation of explicit form of the
Killing vector \cite{Killing, Ibragimov} that corresponds
to the space of independent variables. We associate with higher
order symmetry operators more complex fundamental objects that we
call Killing tensors of rank $j$ and order $s$, with $j,s =1,2,
\ldots$ and conformal Killing tensors of rank $j$ and order $s$.

In this paper we give the definition of the mentioned tensors as a
a complete set of linearly independent solutions of some
overdetermined systems of PDE and find these tensors were found in
explicit form for arbitrary fixed $j$ and $s$ in Minkowsky space
of dimension $p+q \leq 4$. The results can be used in
investigation of higher symmetries of a wide class of systems of
partial differential equations of mathematical physics given in
the same space, in particular, of relativistic and
galilei-invariant wave equations.

Let us describe briefly arrangement of our presentation. Main
definitions related to higher order symmetry operators, are
adduced in Section~1, definition of Killing tensors of rank $j$
and order $s$ is given in Section~2, first order Killing tensors
of rank $j$ and order $s$ in explicit form are found in Sections~3
and~4, conformal Killing tensors and Killing tensors of arbitrary
rank and order are shown in Sections~6,~8 and~9. A complete set of
$n$-th order symmetry operators for equation~(1) with zero and
non-zero ``mass'' $\varkappa$ are adduced in Sections~5 and~7.

\section{Symmetry operators of order $\boldsymbol n$}

For the purpose of our study it is sufficient to consider only
solutions of equation (1) defined on an open set $D$ of the
four-dimensional manifold ${\mathbb R}_{p+q}$ consisting of points with
co-ordinates $(x_1,x_2,\ldots,x_{p+q})$ and analytical with
respect to real variables $x_1,x_2,\ldots,x_{p+q}$. The set of all
such solutions forms a complex vector space that we designate by
the symbol ${\mathcal F}_0$. Setting $D$ as fixed (e.g. assuming that $D$
coincides with ${\mathbb R}_{p+q}$), we will call ${\mathcal F}_0$ the set of
solutions of equation (1).

Let us designate with ${\mathcal F}$ a vector space of all complex-valued
functions defined on $D$ and being real analytical, and with $L$ a
linear differential operator (1) defined on ${\mathcal F}$. Then $L\psi
\in {\mathcal F}$ when $\psi \in {\mathcal F}$. At that ${\mathcal F}_0$ is such subspace of the
vector space ${\mathcal F}$ that coincides with zero-space (kernel) of the
operator $L$.

Let $\mathfrak{M}_n$ be a set (class) of differential operators of
the order $n$ defined on ${\mathcal F}$. Then a symmetry operator $Q \in
\mathfrak{M}_n$ of equation (1) is defined as follows.

\medskip

\noindent
{\bf Definition.} {\it A linear differential operator of order $n$
\begin{gather}
 Q=\sum^n_{i=0} Q_1, \quad Q_i=H^{(a_1 a_2 \ldots a_i)}
\frac{\p^i}{\p x_{a_1} \p x_{a_2} \ldots \p x_{a_i}}, \quad
H^{(a_1 a_2 \ldots a_i)} \in \mathcal{F}
\end{gather}
is called a symmetry operator of  equation {\rm (1)} in the class
$\mathfrak{M}_n$ (or symmetry operator of order $n$) if
\begin{gather}
 [Q, L]=\alpha_Q L, \quad \alpha_Q \in \mathfrak{M}_{n-1},
\end{gather}
 where $[Q, L]=Q L-L Q$ is the commutator of the operators $Q$ and
 $L$.}

\medskip

 Relation (5) should be understood in the sense that operators in
 the right-hand and left-hand parts give the same acting on an
 arbitrary function $\varphi \in {\mathcal F}$. Functions $H^{a_1,a_2,
 \ldots, a_i}$ to be determined are symmetric tensors of rank
 $i$. Hereinafter the parentheses enclose the set of symmetric
 indices.

It is easy to see that the relation (3) follows from (3) for each
$\varphi \in {\mathcal F}_0$. The reverse statement is also true: if an
operator (4) satisfies the relation (3) for arbitrary $\psi \in
{\mathcal F}_0$ then the condition (5) is satisfied for such operator with
some operator $\alpha_Q$. In the case $n=1$ the symmetry operators
defined above may be interpreted as generators of the symmetry
group of the equation being considered \cite{Miller}. We can
show that the set of symmetry operators $Q \in \mathfrak{M}_1$
generates a Lie algebra, and corresponding finite transformations
from the invariance group may be obtained by integration of the
Lie equations \cite{Miller,F3}.

Symmetry operators of order $n > 1$ are not generators of a Lie
algebra anymore and characterize generalized (non-Lie) symmetry of
an equation under study. The problem of description of a complete
set of $n$-th order for equation (1) be reduced to finding of the
general solution of the operator equations (5).

\section{Equations for coefficients of symmetry operators.\\
Killing tensors of rank $\boldsymbol j$ and order $\boldsymbol s$}

For simplification of further calculations it is more convenient
to present the operator $Q$ (4) as the sum of $i$-multiple
anticommutators
\begin{gather}
 Q= \sum^n_{j=0} \hat Q_j,
\end{gather}
where
\begin{gather}
 \hat Q_j=\left[ \ldots \left[ \left[ F^{(a_1 a_2 \ldots a_j)},
\frac{\p}{\p x_{a_1}}\right]_+, \frac{\p}{\p x_{a_2}}\right]_+, \ldots
\frac{\p}{\p x_{a_j}}\right]_+,
\end{gather}
$[A, B]_+=AB+BA$, $F^{(a_1 a_2 \ldots a_i)}$ is a symmetric
tensor of rank $i$. Expanding anticommutators and transferring
differentiation operators to the righthand side, it is possible to
reduce the expression (6) for the operator $Q$ to the form (4),
and, vice versa, to write down any operator of the form (4) as
(6).

We can use a similar representation for the operator $\alpha_Q \in
\mathfrak{M}_{n-1}$
\begin{gather}
 \alpha_Q=\sum^{n-1}_{i=n-2} \hat \alpha_i,
\quad
\hat \alpha_i=\left[ \ldots \left[ \left[ f^{a_1 a_2 \ldots a_i},
\frac{\p}{\p x_{a_1}}\right]_+, \frac{\p}{\p x_{a_2}}\right]_+, \ldots,
\frac{\p}{\p x_{a_i}}\right]_+,
\end{gather}
and write the product $\alpha_Q L$ as
\begin{gather} \alpha_Q L \equiv \frac{1}{4} \left[ \left[ \alpha_Q,
\frac{\p}{\p x_\mu}\right]_+, \frac{\p}{\p x^\mu}\right]_+ +
\frac{1}{2} \left[ \alpha_{Q_\mu}, \frac{\p}{\p x_{\mu}}\right]_+,
\end{gather}
where $\alpha_{Q_\mu}=\frac{\p \alpha_Q}{\p x_\mu}$.

Using the representations (6)--(9) and taking into account that
\begin{gather}
 [Q, L]= \left[Q_\mu, \frac{\p}{\p x^\mu} \right]_+, \quad Q_\mu=\frac{\p Q}{\p x_\mu},
\end{gather}
it is possible to reduce the operator equation (5) to the system
of equations for coefficients $f^{a_1 a_2 \ldots a_i}$ and
$F^{a_1a_2 \ldots a_j}$. In fact, substituting (6)--(10) into (5)
and putting equal coefficients at identical degrees of operators
of differentiation, we obtain
\begin{gather}
 \p^{(a_{j+1}} F^{a_1a_2 \ldots a_j)}=0, \\
f^{(a_1 a_2 \ldots a_i)} \equiv 0 .
\end{gather}
Here $\p^{a_{j+1}}=\frac{\p}{\p x_{a_{j+1}}}$, the round brackets
contain symmetric indices (so symmetrization is implied (11)):
\begin{gather*}
 \frac{1}{j!}\p^{(a_{j+1}}F^{a_1a_2\ldots a_j)}=\p^{a_{j+1}}
F^{a_1a_2 \ldots a_j}+\p^{a_1} F^{a_{j+1}a_2 \ldots a_j}\\
\phantom{\frac{1}{j!}\p^{(a_{j+1}F^{a_1a_2\ldots a_j})}=}{}+
\p^{a_2} F^{a_1 a_{j+1}\ldots a_j} + \cdots + \p^{a_j} F^{a_1a_2
\ldots a_{j+1}},
\end{gather*}
$ F^{a_1a_2 \ldots a_j}$ is a symmetric tensor of the rank $j$.

If $\varkappa = 0$, the equation for coefficients of the symmetry
operator takes the following form:
\begin{gather}
\p^{(a_{j+1}}F^{a_1a_2\ldots a_j)}=\delta^{(a_j a_{j+1}}
f^{a_1a_2\ldots a_{j-1})},
\end{gather}
where  $F^{a_1a_2 \ldots a_j}$ and $f^{a_1 a_2 \ldots a_{j-1}}$
are symmetric tensors with zero trace.

Convoluting equations (13) with respect to one pair of indices, we
can eliminate the unknown functions $f^{a_1 a_2 \ldots
a_{j-1}}$. As a result we get
\begin{gather} \p^{(a_{j+1}}F^{a_1a_2\ldots a_j)}-\frac{j}{m+j-1} \p^b
F^{b(a_2a_3\ldots a_j} g^{a_1a_{j+1})}=0,
\\
f^{a_1a_2\ldots a_{j-1}}=\frac{j}{m+j-1}\p^b F^{b(a_1a_2\ldots a_{j-1}} g^{a_j a_{j+1})},
\end{gather}
where $m=p+q$ is dimension of the space of independent variables.

We see that the problem of description of symmetry operators of
order $n$ for the equation (1) with $\varkappa = 0$ appears to be
equivalent to finding of the general solution of the system of
partial differential equations given by the formula (11). This
system is split with respect to the index $j$, as it splits into
independent subsystems corresponding to $j=0,1, \ldots , n$. As it
will be shown below, for complete description of the symmetry
operators it is actually sufficient to solve only two such
subsystems corresponding to $j=n$ and $j=n-1$.

In the case $j=1$ the system (11) coincides with the Killing
equations \cite{Ibragimov,Killing}, and for $j=2$ it
coincides with equations for the Killing tensor \cite{F3} in the
flat de Sitter space. The corresponding equations (14) determine
conformal Killing vector and Killing tensor in the
$p+q$-dimensional Minkowsky space.

We shall call functions $F^{a_1a_2 \ldots a_j}$ satisfying
equations (11) (or (14)) Killing tensors (or conformal Killing
tensors) or rank $j$ and order 1. The meaning of the term ``order~1''
(that we will omit sometimes) will be explained below.

The equations (11), (13) for a tensor of arbitrary rank were
introduced (in the case $j>2$, without relation to any particular
problem) in the paper \cite{Walker}. However, the general
solution of these equations, as far as we are aware, was obtained
in an explicit form only for $j=1$ and $j=2$ \cite{Katzin}.

In the process of investigation of higher order symmetry operators
admitted by systems of partial differential equations, we have to
deal with more complicated equations for coefficients of such
operators than those given by formulae (11) or (14). These
equations include derivatives of the order $s>1$ and have the form~\cite{F4}
\begin{gather}
\p^{(a_{j+1}} \p^{a_{j+2}} \cdots \p^{a_{j+s}} F^{a_1a_2 \ldots a_j)}=0,
\end{gather}
where $F^{a_1a_2 \ldots a_j}$ is a symmetric tensor, and
\begin{gather}
 \left[ \p^{a_{j+1}}\p^{a_{j+2}} \cdots \p^{a_{j+s}} \tilde F^{a_1a_2 \ldots a_j}
\right]^{SL}=0,
\end{gather}
where $\tilde F^{a_1a_2 \ldots a_j}$ is a symmetric tensor with
zero trace, and the symbol $\left[\cdot  \right]^{SL}$ designates
the zero trace part of the tensor inside the square brackets (in
our case it is a symmetric tensor of the rank $R=j+s$):
\begin{gather} \left[ G^{a_1a_2\ldots a_R}\right]^{SL}=G^{a_1a_2\ldots a_R}+
\sum^{\{ \frac{R}{2}\}}_{d=1} (-1)^d K_d \left(
\prod^d_{i=1}g^{a_{2i-1}a_{2i}} \right) \nonumber\\
\phantom{\left[ G^{a_1a_2\ldots a_R}\right]^{SL}=} {}\times F^{a_{2d+1}
 a_{2d+2\ldots a_i b_1b_2b_3b_4\ldots b_{2d-1}
b_{2d}}}_{}{}_{g_{b_1b_2} g_{b_3b_4\ldots} g_{b_{2d-1}b_{2d}}},
\end{gather}
where $\{ \frac{R}{2}\}$ is the integer part of the number
$\frac{R}{2}$,
\begin{gather} K_d=\frac{n!}{(n-2d)!2^{d-1}}
\prod^d_{i=1} \frac{1}{2(n-i)+m-2},
\end{gather}

In the case $s=1$ the equations (16) and (17) can be reduced to
equations (11) and (14) respectively.

We will call a symmetric tensor $F^{a_1a_2 \ldots a_j}$ satisfying
equations (16) a Killing tensor of rank $j$ and order $s$.We will
call a symmetric tensor $\tilde F^{a_1a_2 \ldots a_j}$ with zero
trace satisfying equations (17) a conformal Killing tensor of rank
$j$ and order $s$.

In Sections 3--7 below we obtain the general solution of equations
(11), (14) for arbitrary $j$ in the space of dimension $p+q \leq
4$. Equations (16), (17) are discussed in Sections 8, 9 where
their general solution is found for $p+q \leq 4$ and arbitrary $j$
and $s$.

\section{Reduction of equations for symmetry operators\\ to a
system of linear algebraic equations}

Let us start investigation of the system of equations (11)
describing the Killing tensor of rank $j$ and order 1.

The system (11) may be written in the following symbolic form:
\begin{gather}
 \mathcal{F}^{a_1a_2\ldots a_{j+1}}=0,
\end{gather}
where $\mathcal{F}^{a_1a_2\ldots a_{j+1}}$ is a symmetric tensor
of rank $j+1$ in $m=p+q$-dimensional space, and unknown functions
are components of symmetric tensor of rank $j$ in $m$-dimensional
space. Whence we can see that the system under investigation is
overdetermined, including $\binom{j+m}{j+1}$ equations for
$\binom{j+m-1}{j}$ unknowns, $\binom{b}{a}=\frac{b!}{a!\left(
b-a\right) !}$ designating binomial coefficients.

Following the general method for solving of overdetermined systems
of partial differential equations \cite{Cartan}, we consider the
set of differential consequences of the system (11), obtained by
differentiation of each term, $k$ times by $x_{b_i}$ $(i=1,2,
\ldots, k)$. For each fixed $k$ such differential consequences are
systems of linear homogeneous algebraic equations for derivatives
\begin{gather} \p^{b_1}  \p^{b_2} \cdots \p^{b_k} \p^{a_{j+1}} F^{a_1a_2\ldots a_j} \equiv
F^{(a_1a_2\ldots a_j, a_{j+1})b_1b_2\ldots b_k} .
\end{gather}
These systems have the form
\begin{gather}
F^{(a_1a_2\ldots a_j, a_{j+1}) b_1b_2\ldots b_k}=0 .
\end{gather}

The system of equations (22) determines condition for vanishing of
the tensor of rank $j+k+1$ symmetric with respect to $j+1$ indices
$a_1,a_2, \ldots, a_{j+1}$ and with respect to $k$ indices
$b_1,b_2, \ldots, b_k$, with unknown components of the tensor (21)
of rank $j+k+1$ symmetric with respect to $j$ indices $a_1,a_2,
\ldots, a_j$ and with respect to $k+1$ indices $a_{j+1},b_1,
\ldots, b_k$. Whence we conclude that the corresponding numbers of
equations $(N_{\mbox{\scriptsize e}})$ and of unknown variables $(N_{\mbox{\scriptsize u}})$ are given by
the formulae
\begin{gather}
 N_{\mbox{\scriptsize e}}=\binom{{j+m}}{{j+1}} \binom{{k+m-1}}{{k}}, \quad
N_{\mbox{\scriptsize u}}=\binom{{j+m-1}}{{j}} \binom{{k+m}}{{k+1}},
\end{gather}
where $m=p+q$ is dimension of the Minkowsky space where equations
(11) are determined (that is the number of independent variables
$x_1,x_2, \ldots$ of the function $F^{a_1a_2\ldots a_j}$).

According to (23)
\begin{gather}
 N_{\mbox{\scriptsize e}} < N_{\mbox{\scriptsize u}}, \quad k < j, \quad
 N_{\mbox{\scriptsize e}} = N_{\mbox{\scriptsize u}} , \quad k=j.
\end{gather}

The formulae (23) allow calculation of the number of linearly
independent solutions of equations (22), as the following
statement is true:

\medskip

\noindent
{\bf Theorem 1.} {\it The system of linear algebraic equations {\rm (22)} is not
degenerate.}

\medskip

Proof of Theorem 1 is adduced below in Appendix.

We conclude from (24) in virtue of Theorem 1 that for $k=j$ the
system of homogeneous linear algebraic equations (22) has only
trivial solutions,
\[
F^{a_1a_2\ldots a_j, a_{j+1} b_1b_2\ldots b_j} \equiv 0.
\]
Whence coefficients of the symmetry operator $F^{a_1a_2\ldots
a_j}$ are polynomials on $x_a$  $(a=1,2, \ldots, m)$ of order $j$.
It follows from (23) that such polynomial contains $N^m_j$
arbitrary parameters, where
\begin{gather}
 N^m_j= \sum^j_{k=0}(N^k_{\mbox{\scriptsize e}}-N^k_{\mbox{\scriptsize u}})=\frac{1}{m}
\binom{{j+m-1}}{{m-1}} \binom{{j+m}}{{m-1}}.
\end{gather}

We see that equations (11) have $N^m_j$ linearly independent
solutions that form a complete system. To find these solutions in
explicit form it is necessary to find the general solution of the
system of linear homogeneous equations (22) for arbitrary given
$j$, $m$ and $k<j$, and then reconstruct polynomials
$F^{a_1a_2\ldots a_j}$ by found values of derivatives of the
tensors $F^{a_1\ldots a_j, a_{j+1} b_1b_2\ldots b_k}$ (let us
remind that indices after the comma designate derivatives with
respect to the corresponding arguments). The general solution of
equations~(11) is adduced in Section 4 below.

\section{Explicit form of Killing tensor of rank $\boldsymbol j$}

According to the above proof, calculation of the explicit form of
Killing tensor of rank $j$ is reduced to finding of the general
solution of non-degenerate system of linear homogeneous algebraic
equations given by the formula (22). Actual solution of this
system with arbitrary given $j$ and $m$ is a rather difficult task
that may be circumvented using the following observation.

\medskip

\noindent
{\bf Lemma 1.} {\it Let $F^{a_1a_2\ldots a_{j_0}}$ be an arbitrary solution
of the system {\rm (11)} for $j=j_0$, and $F^a$ be a solution of the
same system for $j=1$. Then the function
\begin{gather}
 F^{a_1a_2\ldots a_{j_0+1}}=F^{(a_1a_2\ldots a_{j_0}} F^{a_{j_0+1})}
\end{gather}
is a solution of the system {\rm (11)} for $j=j_0 + 1$.}

\medskip

{\bf Proof} is elementary and can be done by direct check.

Lemma 1 given an efficient algorithm for construction of solutions
of equations (11). In fact, solutions of these equations for $j=1$
are well-known: they are Killing tensors
\cite{Ibragimov,Killing}, and a solution for arbitrary
$j$ may be obtained from a solution for $j=1$ by successive
application of the formula (26). If we manage to construct this
way $N^m_j$ linearly independent solutions where $N^m_j$ is given
by the formula (25), then such solutions form a complete system in
virtue of Theorem~1.

Using the algorithm presented above we managed to obtain the
general solution of equations (11) for $m \leq 4$ in the form
\begin{gather}
F^{a_1a_2\ldots a_j}=g^{(a_{j-1}a_j}F^{a_1a_2\ldots a_{j-2})}+f^{a_1a_2\ldots a_j},
\end{gather}
where $F^{a_1a_2\ldots a_{j-2}}$ is the general solution of
equations (11) for $j \to (j-2)$ depending on $N^m_{j-2}$
arbitrary parameters, and $f^{a_1a_2\ldots a_{j}}$
is a solution of
equations (11) depending on $N_j^m-N^m_{j-2}$
arbitrary parameters.

The first addend in the right-hand part of the formula (27)
corresponds to such symmetry operator (7) of order $j$ that on the
set of solutions of equation (1) can be reduced to a symmetry
operator of order $j-2$. Explicit expressions for $f^{a_1a_2\ldots
a_j}$ corresponding to $m \leq 4$ are adduced below.

1. $m=1$. The corresponding tensor $f^{a_1a_2\ldots a_j}$ can be
reduced to a scalar not depending on the only variable.

2. $m=2$. Tensors $f^{a_1a_2\ldots a_j}$ depend on two variables
$x_1$ and $x_2$. The number of independent solutions, according to
(25), is
\begin{gather}
N=N^2_j-N^2_{j-2}=2j+1.
\end{gather}
Solutions are numbered by an integer number $c$ satisfying the
condition
\begin{gather}
0 \leq c \leq j,
\end{gather}
and include for $c=0$ one, and for each $c>0$ two arbitrary
parameters giving independent components of a symmetric zero trace
tensor $\lambda^{ a_1a_2\ldots a_{j-c}}$ of rank $j-c$. The
explicit form of the corresponding solution $f^{a_1a_2\ldots
a_j}_c$ is given by the formula
\begin{gather}
f^{a_1a_2\ldots a_j}_c=\varepsilon \hat f^{a_1a_2\ldots a_j}+(1-\varepsilon)
\hat f^{(a_1a_2\ldots a_{j-1}}\varepsilon^{a^j)b}x_b,
\end{gather}
where $\varepsilon^{a^jb}$ is the unit antisymmetric tensor,
$\varepsilon=\frac{1}{2} [1+(-1)^c]$,
\begin{gather}
\hat f^{a_1a_2\ldots a_j} =\lambda^{( a_1a_2\ldots a_{j-c}}
\sum^{\{\frac{c}{2}\}}_{\mu=0} \left( \prod^{j-c+2\mu}_{i=j-c+1}
x^{a_i}\right)^* \nonumber\\
\phantom{\hat f^{a_1a_2\ldots a_j} =}{}\times \left( \prod^{\min \{ \frac{j}{2},
\frac{j+1}{2}-l\}}_{k=\{\frac{j-c}{2}\}+\mu+1}
g^{a_{2k+l}a_{2k})}\right)^*(-1)^\mu \binom{\{\frac{c}{2}\}}{\mu} (x^2)^{\{\frac{c}{2}\}-\mu},\nonumber\\
 \left( \prod^B_{\lambda=A} f_\lambda\right)^*=\left\{
\begin{array}{ll} \displaystyle \prod^B_{\lambda=A} f_\lambda, &  B \geq A,\vspace{1mm}\\
1,  &  B < A,\end{array} \right.  \quad x^2=x^2_1+x^2_2,
 \quad l=(-1)^{j+c+1},
\end{gather}
and symmetrization over the indices $a_1,a_2,\ldots, a_j$ is
implied.

3. $m=3$. The tensor $f^{a_1a_2\ldots a_j}$ depends on three
variables $\vec{x}=(x_1,x_2,x_3)$. The number of
independent solutions is equal to
\begin{gather}
N=N^3_j-N^3_{j-2}=\frac{1}{3}(j+1)(2j^2+4j+3).
\end{gather}

The solutions are numbered with pairs of integers  $c=(c_1,c_2)$
satisfying the conditions
\begin{gather}
0 \leq c_1 \leq 2 \left\{\frac{j}{2}\right\}, \quad 0 \leq c_2 \leq j-i
\left\{\frac{c_1+1}{2}\right\}, \quad \varepsilon_a=\frac{1}{2}[1+(-1)^a],
\end{gather}
and include for each $c$ the set $2c_1+1$ of arbitrary parameters
giving independent components of a symmetric zero trace tensor
$\lambda^{a_1a_2\ldots a_{c_1}}$ of rank $c_1$. Explicit forms of
the corresponding solutions $f^{a_1a_2\ldots a_j}_c$ are given by
the formula
\begin{gather}
f_c^{a_1a_2\ldots a_j}=\varepsilon_{c_2} \hat f_{c_1c_2}^{a_1a_2\ldots a_j}+
(1-\varepsilon_{c_2}) \hat f_{c_1c_2}^{b(a_1a_2\ldots a_{j-1}}\varepsilon^{a_j)bc}x_c,
\end{gather}
where $\varepsilon^{a^jbc}$ is the unit antisymmetric tensor,
\begin{gather}
\hat f_{c_1c_2}^{a_1a_2\ldots a_j}=\sum_\mu K_\mu
\lambda^{\beta_\mu, (A_\mu}_c \left(
\prod^{A_\mu+L_\mu}_{i=A_\mu+1} x^{a_i}\right)^* \left(
\prod^{\min (\{ \frac{j}{2}\},
\{\frac{j+1}{2}\}-l)}_{k=\{\frac{1}{2} (A_\mu+L_\mu)\}+1}
g^{a_{2k}a_{2k+l}} \right)^* (x^2)^{F_\mu}.
\end{gather}
Here
\begin{gather}
\mu=( \mu_1, \mu_2, \mu_3, \mu_4, \mu_5), \quad x^2=x_a x_b g^{ab},
\nonumber\\
K_\mu=(-1)^{\mu_1+\mu_3+\mu_5} 2\mu_3 \frac{ \{
\frac{c_2}{2}\}!}{\mu_2! \mu_3! \mu_4!} \binom{\{
\frac{c_1}{2}\}}{\mu_1},\nonumber\\
B_\mu=2 \mu_2+\mu_3+\mu_5, \quad A_\mu=j-c_1-B_\mu, \quad l=(-1)^{c_2+j+1},
\nonumber\\
L_\mu=c_1+\mu_3-2\mu_1-\mu_5, \quad F_\mu=\mu_1+\mu_4,
\end{gather}
and $\lambda^{B_\mu, A_\mu}$ is an arbitrary symmetric zero trace
tensor of rank $A_\mu+B_\mu$ convoluted with $B_\mu$ vectors
$x_k$:
\begin{gather}
\lambda^{B_\mu, A_\mu}=\lambda^{b_1b_2 \ldots b_{B_\mu} a_1a_2 \ldots a_{A_\mu}}
x_{b_1}x_{b_2}\cdots x_{b_{B_\mu}} .
\end{gather}
Summation in (35) is to be done over all possible nonnegative
values of $\mu$ satisfying the conditions
\begin{gather}
0 \leq \mu_1 \leq \left\{ \frac{c_1}{2}\right\}, \quad \mu_2+\mu_3+\mu_4=\left \{ \frac{c_2}{2}\right\}, \quad
0 \leq \mu_5 \leq \frac{1}{2} [1-(-1)^{c_1}].
\end{gather}

As well as in the formula (31) symmetrization over the indices
$a_1,a_2,\ldots, a_j$ in the right-hand part of (35) is implied.

4. $m=4$. The tensor $f^{a_1a_2\ldots a_j}$ depends on four
variables $x=(x_1,x_2,x_3,x_4)$. The number of independent
solutions is equal to
\begin{gather}
 N=N^4_j-N^4_{j-2}=\frac{1}{4!}(j+1)(j+2)(2j+3)(j^2+3j+4).
\end{gather}

Solutions are numbered with triples of integers $c=(c_1,c_2,c_3)$
satisfying conditions (33) and (40):
\begin{gather}
 0 \leq c_3 \leq j-2 \left\{ \frac{c_1+1}{2}\right\} - 2 c_2,
\end{gather}
and include for each $c$ a set of $N_c$ arbitrary parameters,
where
\begin{gather}
 N_c=\left\{ \begin{array}{ll}
(c_2+2c_3+1)^2, & c_1=c_2+2 c_3 \\
2(c_2+2c_3+1)(2c_1-c_2-2c_3+1), & c_1 \not= c_2+2c_3, \end{array} \right.
\end{gather}
These parameters give independent components of an irreducible
tensor
\[
\lambda^{a_1a_2\ldots a_{R_1} [a_{R_1+1}b_1][a_{R_1+2}b_2]
\ldots [a_{R_1+R_2} b_{R_2}]},
\] where $R_1=c_2+2c_3$,
$R_2=c_1-c_2-2c_3$ (let us remind that an irreducible tensor of rank
$R_1+2R_2$ has $R_1$ symmetric indices and $R_2$ symmetric pairs
of antisymmetric indices, and convolution by any pair of indices
and any triple of indices with completely antisymmetric tensor
$\varepsilon_{\mu \nu \rho \sigma}$ vanishes). Explicit expressions
for the respective solutions are given by the formula (42):
\begin{gather}
 f_{c}^{a_1a_2\ldots a_j}= \sum_\mu K_\mu \lambda^{\beta_\mu, (A_\mu, D_c}
\left( \prod^{A_\mu+D_c+L_\mu}_{i=A_\mu+D_c+1} x^{a_i}\right)^*\nonumber\\
\phantom{ f_{c}^{a_1a_2\ldots a_j}=}{}\times
\left( \prod^{ \{ \frac{j}{2}\}}_{k=\{\frac{1}{2} (A_\mu+D_c+L_\mu)\}+1}
g^{a_{2k+l}a_{2k})} \right)^* (x^2)^{F_\mu},
\end{gather}
where $\mu$, $x^2$,  $K_\mu$, $B_\mu$, $F_\mu$ are given by the
formulae (36), (38)
$a,b=1,2,3,4$, $A_\mu=j-c_1-B_\mu-D_c$, $D_c=c_3$, $l=(-1)^{n+1}$,
\begin{gather}
\lambda^{B_\mu, A_\mu, D_c}=\lambda^{b_1b_2 \ldots b_{B_\mu}
a_1a_2\ldots a_{A_\mu} [a_{A_\mu+1}d_1] \ldots [a_{A_\mu+D_c}
d_{D_c}]} \nonumber\\
\phantom{\lambda^{B_\mu, A_\mu, D_c}=}{}
 \times x_{b_1} x_{b_2}\cdots x_{b_{B_\mu}}x_{d_1} x_{d_2} \cdots x_{d_{D_c}},
\end{gather}
symmetrization over the indices $a_1, a_2, \ldots, a_j$ is implied
in the right-hand side of (42).

So, we have obtained the general solution of equations (11) in the
space of dimension $ m \leq 4$. One can verify by a direct check
that the found solutions satisfy equations (11) and are linearly
independent (it is not difficult to prove the latter considering
$i$-fold convolutions of the found solutions with $g^{kl}, 0 \leq
i \leq \{ \frac{j}{2}\}$). On the other side, these solutions form
a complete system, as the number of arbitrary parameters they
include is in compliance with the formula (25).

Let us also mention that we can present the general solution of
equations (11) also in the form
\begin{gather}
F^{a_1a_2\ldots a_j}= \sum^j_{l=0} \lambda^{a_1a_2\ldots
a_l [a_{l+1}b_1][a_{l+2}b_2] \ldots [a_j b_{j-l}]} x_{b_1} x_{b_2}
\ldots x_{b_{j_l}},
\end{gather}
where $\lambda^{a_1a_2\ldots a_l [a_{l+1}b_1] \ldots
[a_j b_{j-l}]}$ is a tensor, symmetric with respect to permutation
of indices $a_1,  \ldots, a_j$ and antisymmetric with respect
to permutation of indices $a_{l+j}$ with $b_i$, $1 \leq i \leq
j-l$, with vanishing convolution of this tensor over any three
indices with $\varepsilon_{\mu \nu \rho \sigma}$. The latter means
that a cyclic permutation with respect to any triple of indices
$(a_k, a_{l+s}, b_s)$ gives zero, so the polynomial (44)
admittedly satisfies equation (11). On the other side, the number
of independent components of the tensor $\lambda^{a_1a_2\ldots a_l
[a_{l+1}b_1] \ldots [a_j b_{j-l}]}$ for $0 \leq l
\leq j$ is exactly $N_j^m$ (25), so the formula (44) gives the
general solution of equations (11). Decomposing tensors
$\lambda^{a_1a_2\ldots a_l [a_{l+1}b_1] \ldots [a_j
b_{j-l}]}$, $0 \leq l \leq j$ into irreducible ones (that is
having vanishing convolutions with respect to any pair of
indices), we come to formulae (26)--(43).

We formulate the above results as the following theorem.

\medskip

\noindent
{\bf Theorem 2.} {\it Equations {\rm (11)} in a space of dimension $m \leq 4$ have
$N_j^m$ linearly independent solutions. These solutions are
polynomials of $x_a$ of degree $j$ and are given in explicit form
by relations {\rm (26)--(43)}.}

\medskip

The above theorem determines the explicit form of the Killing
tensor of rank $j$ in a space of dimension $m \leq 4$.

\section{Explicit form of symmetry operators $\boldsymbol{Q_n}$ for $\boldsymbol{n \leq 4}$}

The above results allow presenting in explicit form of symmetry
operators of order $n$ for equation (1) in $m$-dimensional space
for arbitrary given $n<\infty$ and $m \leq 4$. For this purpose it
is sufficient to look through all admissible values of $c$ given
by the formulae (29), (33), (40) and construct in accordance to
the formulae (26)--(43) the corresponding expressions for Killing
tensors of rank $j$ $F^{a_1a_2 \ldots a_j}_c$ (following (27), it is
sufficient to restrict oneself with construction of $f^{a_1a_2 \ldots
a_j}$), then substitute obtained expressions into (6), (7) and sum
up over $j$ from $0$ to $n$.

In this Section we will realize this program for all $n\leq 4$ and
$m \leq 4$, and write down in explicit form the corresponding
symmetry operators.

Let us calculate the number of linearly independent symmetry
operators of order $n$. It is equal, according to (27) to the
number of linearly independent solutions of the system (11) for
$j=n, n-1$ or (see (25))
\begin{gather}
 N(n, m)=N^n_m+N^{n-1}_m= \frac{2n^2+2mn+m(m-1)}{m(m-1)}
\binom{n+m-2}{m-2} \binom{n+m-1}{m-2},\!\!\!
\end{gather}

In particular, for $m=2,3,4$
\begin{gather} N(n,2)=(n+1)^2, \nonumber\\
N(n,3)=\frac{1}{6} (n+1)(n+2)(n^2+3n+3),\nonumber\\
N(n,4)=\frac{1}{72} (n+1)(n+2)^2(n+3)(n^2+4n+6).
\end{gather}
Values of these numbers for $n=1,2,3,4$ are adduced in Table 1.

\newpage

\centerline{\small {\bf Table 1.} Number of symmetry operators of order $n$}

\centerline{\small for equation (1) in $m$-dimensional space.}

\medskip

\centerline{\begin{tabular}{|c|cccc|}
\hline
$m\Big\backslash n$  &  $1$   &  $2$ &  $3$  &  $4$  \\
\hline
$2$    &  $4$   &  $9$   &  $16$ &  $25$  \\
$3$    &  $7$   &  $26$   &  $70$ &  $155$  \\
$4$    &  $11$   &  $60$   &  $225$ &  $665$ \\
\hline
\end{tabular}}
\medskip

Let us write down explicitly the corresponding solutions
$F^{(n)}_m=F^{a_1 \ldots a_n}_m$ of equa\-tions~(11)

$m=1$
\[ n=1,2,3,4, \quad F^{(n)}_1=\lambda_{1n}; \]

$m=2$
\begin{gather}
n=0, \quad F^{(0)}_2=\lambda_{20};\nonumber
\\
n=1, \quad F^a_2=\lambda^a_{21}+\lambda_{21} \varepsilon^{ab} x_b;\nonumber\\
n=2, \quad F^{a_1a_2}_2=g^{a_1a_2} F^{(0)}_2+\lambda^{a_1a_2}+\lambda^{(a_1}\varepsilon^{a_2)^b} x_b
+\lambda(g^{a_1a_2}x^2-x^{a_1} x^{a_2});
\\
n=3, \quad F^{a_1a_2a_3}_2=g^{(a_1a_2} F^{a_3)}_2+\lambda^{a_1a_2a_3}_{(0,0)}
+\lambda^{(a_1a_2}_{(0,1)} \varepsilon^{a_3)^b} x_b
\nonumber\\
\phantom{n=3, \quad}{}+\lambda^{(a_1}_{(0,2)}(g^{a_2a_3)}x^2-x^{a_2}x^{a_3)})+\lambda_{(0,3)}
x^{(a_1} x^{a_2} \varepsilon^{a_3)^b}x_b;
\\
n=4, \quad
 F^{a_1a_2a_3a_4}_2=g^{(a_1a_2} F^{a_3a_4)}_2+
 \lambda^{a_1a_2a_3a_4}_{(0,0)}
+\lambda^{(a_1a_2a_3}_{(0,1)} \varepsilon^{a_4)^b} x_b\nonumber\\
\phantom{n=4, \quad}{}+\lambda^{(a_1a_2}_{(0,2)}(x^{a_3} x^{a_4)} -g^{a_3a_4)}x^2)+
\lambda_{(0,4)} (x^{(a_1} x^{a_2} x^{a_3}
x^{a_4)}\nonumber\\
\phantom{n=4, \quad}{}-2x^{(a_1}x^{a_2} g^{a_3a_4)}+x^4g^{(a_1a_2} g^{a_3a_4)});
\end{gather}

$m=3$
\begin{gather}
n=0, \quad F^{(0)}_3=\lambda_3;
\nonumber\\
n=1, \quad F^a_3=\lambda^a+\lambda^b \varepsilon^{abc}x_c;
\nonumber\\
 n=2, \quad
F^{a_1a_2}_3=g^{a_1a_2}F^{(0)}_3+\lambda^{a_1a_2}_{(0,0)}+
\lambda^{b(a_1}_{(0,1)} \varepsilon^{a_2)bc}x_c+
\lambda^{a_1a_2}_{(0,2)}x^2-2\lambda^{b_1(a_1}_{(0,2)}x^{a_2)}x_{b_1}\nonumber\\
\phantom{ n=2, \quad}{}+\lambda^{b_1b_2}_{(0,2)}x_{b_1} x_{b_2}g^{a_1a_2}+
\lambda^{(a_1}_{(1,0)}x^{a_2)}-\lambda^{b_1}_{(1,0)}x_{b_1}
g^{a_1a_2}+\lambda_{(2,0)}x^{a_1}x^{a_2}
\nonumber\\
\phantom{n=2, \quad}{}-\lambda_{(2,0)}g^{a_1a_2}x^2;\\
n=3, \quad
F^{a_1a_2a_3}_3=g^{(a_1a_2}F^{a_3)}_3+\lambda^{a_1a_2a_3}_{(0,0)}+
\lambda^{b(a_1a_2}_{(0,1)} \varepsilon^{a_3)bc}x_c+
\lambda^{a_1a_2a_3}_{(0,2)}x^2-2\lambda^{b(a_1a_2}_{(0,2)}x^{a_3)}
x_{b}\nonumber\\
\phantom{n=3, \quad}{}
+ \lambda^{b_1b_2(a_1}_{(0,2)}g^{a_2a_3)}x_{b_1} x_{b_2}+
\lambda^{b(a_1a_2}_{(0,3)}\varepsilon^{a_3)bc} x_cx^2-2
\lambda^{b d(a_1}_{(0,3)}x^{a_2}\varepsilon^{a_3)bc} x_c x_d
\nonumber\\
\phantom{n=3, \quad}{}
+\lambda_{(0,3)}^{d b_1b_2}g^{(a_1a_2}\varepsilon^{a_3)d c} x_c
x_{b_1}x_{b_2}+\lambda_{(1,0)}^{(a_1a_2} x^{a_3)}-
\lambda^{b(a_1}_{(1,0)}g^{a_2a_3)}x_b+\lambda^{b(a_1}_{(1,1)}
x^{a_2}\varepsilon^{a_3)bc}x_c\nonumber\\
\phantom{n=3, \quad}{}
+\lambda^{(a_1}_{(2,0)}x^{a_2}x^{a_3)}-
\lambda^{b d}_{(1,1)}g^{(a_1a_2}\varepsilon^{a_3)bc}x_c x_d+
\lambda^b_{(2,1)} x^{(a_1} x^{a_2}
\varepsilon^{a_3)bc}x_c\nonumber\\
\phantom{n=3, \quad}{}
-\lambda^{(a_1}_{(2,0)} g^{a_2a_3)} x^2- \lambda^b_{(2,1)} g^{(a_1a_2}\varepsilon^{a_3)bc} x_c x^2;
\nonumber\\
n=4, \quad F^{a_1a_2a_3a_4}_3=g^{(a_1a_2} F_3^{a_3a_4)}+
\lambda^{a_1a_2a_3a_4}_{(0,0)}+\lambda^{a_1a_2a_3a_4}_{(0,2)}x^2+
\lambda^{b(a_1a_2a_3}_{(0,1)}\varepsilon^{a_4)bc}x_c\nonumber\\
\phantom{n=4, \quad}{}-2\lambda^{b(a_1a_2a_3}_{(0,2)}x^{a_4)}x_b
+\lambda^{b_1b_2(a_1a_2}_{(0,2)} g^{a_3a_4)} x_{b_1}x_{b_2}+
\lambda^{b(a_1a_2a_3}_{(0,3)}\varepsilon^{a_4)bc} x_c x^2\nonumber\\
\phantom{n=4, \quad}{}
-2\lambda^{b d(a_1a_2}_{(0,3)} x^{a_3} \varepsilon^{a_4)bc}x_c
x_d+\lambda^{a_1a_2a_3a_4}_{(0,4)} x^4
+\lambda^{d b_1b_2(a_1}_{(0,3)} g^{a_3a_2}\varepsilon^{a_4)d c}
x_c x_{b_1} x_{b_2}\nonumber\\
\phantom{n=4, \quad}{}
-4 \lambda^{b(a_1a_2a_3}_{(0,4)} x^{a_4)}x_bx^2
+4\lambda^{b_1b_2(a_1a_2}_{(0,4)} x^{a_3} x^{a_4)} x_{b_1}
x_{b_2}+ 2 \lambda^{b_1b_2(a_1a_2}_{(0,4)} g^{a_3a_4)} x_{b_1}
x_{b_2} x^2\nonumber\\
\phantom{n=4, \quad}{}
-4\lambda^{b_1b_2b_3(a_1}_{(0,4)} x^{a_2} g^{a_3a_4)} x_{b_1}
x_{b_2}x_{b_3}+\lambda^{(a_1a_2a_3}_{(1,0)} x^{a_4)}\nonumber\\
\phantom{n=4, \quad}{}
+ \lambda^{b_1b_2b_3b_4}_{(0,4)} x_{b_1}x_{b_2}x_{b_3}x_{b_4}
g^{(a_1a_2}g^{a_3a_4)}
-\lambda^{b(a_1a_2}_{(1,0)}g^{a_3a_4)}x_b+\lambda^{b(a_1a_2}_{(1,1)}
x^{a_3}\varepsilon^{a_4)bc}x_c \nonumber\\
\phantom{n=4, \quad}{}
-\lambda^{b d(a_1}_{(1,1)} g^{a_2a_3}\varepsilon^{a_4)bc} x_c
x_d+\lambda^{(a_1a_2a_3}_{(1,2)}x^{a_4)}x^2
-\lambda^{b(a_1a_2}_{(1,2)}g^{a_3a_4)}x_bx^2\nonumber\\
\phantom{n=4, \quad}{}
-2\lambda^{b(a_1a_2}_{(1,2)}
x^{a_3}x^{a_4)}x_b +
3\lambda^{b_1b_2(a_1}_{(1,2)} x^{a_2} g^{a_3a_4)} x_{b_1}
x_{b_2}+\lambda^{(a_1a_2}_{(2,0)} x^{a_3} x^{a_4)}\nonumber\\
\phantom{n=4, \quad}{}
-\lambda^{b_1b_2b_3}_{(1,2)}x_{b_1}x_{b_2}x_{b_3}g^{(a_1a_2}
g^{a_3a_4)}-\lambda^{(a_1a_2}_{(2,0)} g^{a_3a_4)}x^2+
\lambda^{b(a_1}_{(2,1)} x^{a_2} x^{a_3} \varepsilon^{a_4)bc}x_c\nonumber\\
\phantom{n=4, \quad}{}
- \lambda^{b(a_1}_{(2,1)} g^{a_2a_3}\varepsilon^{a_4)bc}x_cx^2+
\lambda^{(a_1a_2}_{(2,2)} x^{a_3} x^{a_4)} x^2-2
\lambda^{b(a_1}_{(2,2)} x^{a_2} x^{a_3} x^{a_4)} x_b\nonumber\\
\phantom{n=4, \quad}{}
+\lambda^{b_1b_2}_{(2,2)} x_{b_1} x_{b_2} x^{(a_1} x^{a_2}
g^{a_3a_4)}-\lambda^{(a_1a_2}_{(2,2)} g^{a_3 a_4)} x^4+
2\lambda^{b(a_1}_{(2,2)} x^{a_2} g^{a_3a_4)}x_bx^2
\nonumber\\
\phantom{n=4, \quad}{}+\lambda^{(a_1}_{(3,0)} x^{a_2} x^{a_3} x^{a_4)}-
\lambda^{b_1b_2}_{(2,2)} x_{b_1} x_{b_2} x^2
g^{(a_1a_2}g^{a_3a_4)}-\lambda^{(a_1}_{(3,0)}x^{a_2} g^{a_3 a_4)}x^2
\nonumber\\
\phantom{n=4, \quad}{}
-\lambda^{b_1}_{(3,0)} x_{b_1} x^{(a_1}x^{a_2} g^{a_3a_4)}+
\lambda^{b_1}_{(3,0)} x_{b_1} x^2 g^{(a_1a_2} g^{a_3a_4)}+
\lambda_{(4,0)} x^{(a_1}x^{a_2} x^{a_3} x^{a_4)}\nonumber\\
\phantom{n=4, \quad}{}
-2 \lambda_{(4,0)} x^{(a_1}x^{a_2} g^{a_3a_4)}x^2 +
\lambda_{(4,0)} x^4 g^{(a_1a_2} g^{a_3a_4)};
\end{gather}

$ m=4 $
\begin{gather}
n=0, \quad F^{(0)}_4=\lambda;\nonumber\\
n=1, \quad
F^a_4=\lambda^a_{(0,0,0)}+\lambda^{[ad_1]}_{(0,0,1)}x_{d_1};\nonumber\\
n=2, \quad F_4^{a_1a_2}=g^{a_1a_2}
F^{(0)}_4+\lambda^{a_1a_2}_{(0,0,0)}+
\lambda^{(a_1[a_2d_1])}_{(0,0,1)}x_{d_1}
+\lambda^{[a_1d_1][a_2d_2]}_{(0,0,2)}x_{d_1} x_{d_2}\nonumber\\
\phantom{n=2, \quad}+
\lambda^{a_1a_2}_{(0,1,0)}x^2
-2\lambda^{b(a_1}_{(0,1,0)}x^{a_2)}x_b+ \lambda^{b_1b_2}_{(0,1,0)}
x_{b_1}x_{b_2} g^{a_1a_2}+
\lambda^{(a_1}_{(1,0,0)}x^{a_2)}
\nonumber\\
\phantom{n=2, \quad}
-\lambda^{b}_{(1,0,0)} x_{b}
g^{a_1a_2}+\lambda_{(2,0,0)}x^{a_1}x^{a_2}- \lambda_{(2,0,0)}
 g^{a_1a_2} x^2;
\\
n=3, \quad F_4^{a_1a_2a_3}=g^{(a_1a_2}
F^{a_3)}_4+\lambda^{a_1a_2a_3}_{(0,0,0)}+
\lambda^{(a_1a_2[a_3d_1])}_{(0,0,1)}x_{d_1}
+\lambda^{(a_1[a_2d_1][a_3d_2])}_{(0,0,2)}x_{d_1}x_{d_2}\nonumber\\
\phantom{n=3, \quad}
+\lambda^{[a_1d_1][a_2d_2][a_3d_3]} _{(0,0,3)}
x_{d_1} x_{d_2} x_{d_3}-2\lambda^{b_1(a_1a_2}_{(0,1,0)}
x^{a_3)}x_{b_1}+\lambda^{a_1a_2a_3}_{(0,1,0)}x^2
\nonumber\\
\phantom{n=3, \quad}
+ \lambda^{b_1b_2(a_1}_{(0,1,0)}g^{a_2a_3)}x_{b_1}x_{b_2}+
\lambda^{(a_1a_2[a_3d_1])}_{(0,1,1)}x^2 x_{d_1}-
2\lambda^{b_1(a_1[a_2d_1]}_{(0,1,1)}x^{a_3)}x_{b_1}x_{d_1}
\nonumber\\
\phantom{n=3, \quad}
+ \lambda^{b_1b_2([a_1d_1]}_{(0,1,1)} g^{a_2a_3)} x_{b_1} x_{b_2}
x_{d_1}+ \lambda^{(a_1a_2}_{(1,0,0)} x^{a_3)}-
\lambda^{b_1(a_1}_{(1,0,0)}g^{a_2a_3)}x_{b_1}+
\lambda^{(a_1[a_2d_1]}_{(1,0,1)} x^{a_3)}x_{d_1}
\nonumber\\
\phantom{n=3, \quad}
-\lambda^{b_1([a_1d_1]}_{(1,0,1)}g^{a_2a_3)} x_{b_1} x_{d_1}+
\lambda^{(a_1}_{(2,0,0)} x^{a_2}x^{a_3)}-
\lambda^{(a_1}_{(2,0,0)}g^{a_2a_3)} x^2+
\lambda^{([a_1d_1]}_{(2,0,1)} x^{a_2}x^{a_3)}x_{d_1}
\nonumber\\
\phantom{n=3, \quad}
-\lambda^{([a_1d_1]}_{(2,0,1)} g^{a_2a_3)} x^2 x_{d_1};
\\
n=4, \quad F^{a_1a_2a_3a_4}_4=g^{(a_1a_2} F^{a_3a_4)}+
\lambda^{a_1a_2a_3a_4}_{(0,0,0)}+
\lambda^{(a_1a_2a_3[a_4d_1])}_{(0,0,1)} x_{d_1}\nonumber\\
\phantom{n=4, \quad}{}
+\lambda^{(a_1a_2[a_3d_1][a_4d_2])}_{(0,0,2)}x_{d_1} x_{d_2}
+
\lambda^{(a_1[a_2d_1][a_3d_2][a_4d_3])}_{(0,0,3)}
 x_{d_1} x_{d_2}
x_{d_3}\nonumber\\
\phantom{n=4, \quad}{}+
\lambda^{([a_1d_1][a_2d_2][a_3d_3][a_4d_4])} x_{d_1}
x_{d_2} x_{d_3} x_{d_4}
+\lambda^{a_1a_2a_3a_4}_{(0,1,0)}x^2-2
\lambda^{b_1(a_1a_2a_3}_{(0,1,0)} x^{a_4)} x_{b_1}\nonumber\\
\phantom{n=4, \quad}{}+
\lambda^{b_1b_2(a_1a_2}_{(0,1,0)} g^{a_3a_4)}x_{b_1}x_{b_2}
+
\lambda^{(a_1a_2a_3[a_4 d_1])}_{(0,1,1)} x^2 x_{d_1}
-2\lambda^{b_1(a_1a_2[a_3d_1]}_{(0,1,1)}x^{a_4)} x_{b_1} x_{d_1}
\nonumber\\
\phantom{n=4, \quad}{}+
\lambda^{b_1b_2(a_1[a_2d_1]}_{(0,1,1)} g^{a_3a_4)}
 x_{b_1} x_{b_2} x_{d_1}+
\lambda^{(a_1a_2[a_3d_1][a_4d_2])}_{(0,1,2)}x^2 x_{d_1} x_{d_2}
\nonumber\\
\phantom{n=4, \quad}{}-
2\lambda^{b_1(a_1[a_2d_1][a_3d_2]} x^{a_4)} x_{b_1} x_{d_1}
x_{d_2}
+\lambda^{b_1b_2([a_1d_1][a_2d_2]}g^{a_3a_4)} x_{b_1}x_{b_2}
 x_{d_1}x_{d_2}
\nonumber\\
\phantom{n=4, \quad}{}
+\lambda^{a_1a_2a_3a_4}_{(0,2,0)}x^4 -
4\lambda^{b_1(a_1a_2a_3}_{(0,2,0)}x^{a_4)} x^2x_{b_1}+4
\lambda^{b_1b_2(a_1a_2}_{(0,2,0)} x^{a_3}x^{a_4)} x_{b_1}
x_{b_2}\nonumber\\
\phantom{n=4, \quad}{}
+2\lambda^{b_1b_2(a_1a_2}_{(0,2,0)}g^{a_3a_4)}x^2 x_{b_1} x_{b_2}-4
\lambda^{b_1b_2b_3(a_1}_{(0,2,0)} x^{a_2}g^{a_3a_4)}
 x_{b_1} x_{b_2} x_{b_3}
\nonumber\\
\phantom{n=4, \quad}{}
 +g^{(a_1a_2}g^{a_3a_4)}
\lambda^{b_1b_2b_3b_4}_{(0,2,0)}x_{b_1} x_{b_2}x_{b_3} x_{b_4}+
\lambda^{a_1a_2a_3}_{(1,0,0)} x^{a_4)}-
\lambda^{b_1a_1a_2}_{(1,0,0)} g^{a_3a_4)}x_{b_1}
\nonumber\\
\phantom{n=4, \quad}{}
+ \lambda^{(a_1a_2[a_3d_1]}_{(1,0,1)} x^{a_4)} x_{d_1} -
\lambda^{b_1(a_1[a_2d_2]}_{(1,0,1)} g^{a_3a_4)} x_{b_1} x_{d_1}+
\lambda^{(a_1[a_2d_1][a_3d_2]}_{(1,0,2)} x^{a_4)}x_{d_1} x_{d_2}
\nonumber\\
\phantom{n=4, \quad}{}
-\lambda^{b_1(a_1d_1][a_2d_2]}_{(1,0,2)} g^{a_3a_4)} x_{b_1}
x_{d_1} x_{d_2}
+\lambda^{(a_1a_2a_3}_{(1,1,0)}x^{a_4)} x^2-
\lambda^{b_1(a_1a_2}_{(1,1,0)}g^{a_3a_4)} x_{b_1} x^2
\nonumber\\
\phantom{n=4, \quad}{}
-2\lambda^{b_1(a_1a_2}_{(1,1,0)}x^{a_4} x^{a_4)} x_{b_1}+3
\lambda^{b_1b_2(a_1}_{(1,1,0)}x^{a_2} g^{a_3a_4)}x_{b_1} x_{b_2}
-\lambda^{b_1b_2b_3}_{(1,1,0)}g^{(a_1a_2} g^{a_3a_4)} x_{b_1}
x_{b_2} x_{b_3}
\nonumber\\
\phantom{n=4, \quad}{}
+ \lambda^{(a_1a_2}_{(2,0,0)}x^{a_3} x^{a_4}-
\lambda^{(a_1a_2}_{(2,0,0)} g^{a_3a_4)}x^2+
\lambda^{(a_1[a_2d_1]}_{(2,0,1)}x^{a_3}x^{a_4)} x_{d_1}-
\lambda^{(a_1[a_2d_1]}_{(2,0,1)} g^{a_3a_4)}x^2x_{d_1}
\nonumber\\
\phantom{n=4, \quad}{}
+\lambda^{([a_1d_1][a_2d_2]}_{(2,0,2)}x^{a_3}x^{a_4)}  x_{d_1} x_{d_2}-
\lambda^{([a_1d_1][a_2d_2]}_{(2,0,2)}g^{a_3a_4)}x^2  x_{d_1} x_{d_2}+
\lambda^{(a_1a_2}_{(2,1,0)} x^{a_3} x^{a_4)} x^2
\nonumber\\
\phantom{n=4, \quad}{}
-2 \lambda^{(b_1(a_1}_{(2,1,0)}  x^{a_2} x^{a_3} x^{a_4)} x_{b_1}+
\lambda^{b_1b_2}_{(2,1,0)} x^{(a_1} x^{a_2} g^{a_3a_4)} x_{b_1} x_{b_2}
- \lambda^{(a_1a_2}_{(2,1,0)} g^{a_3a_4)} x^4
\nonumber\\
\phantom{n=4, \quad}{}
+2 \lambda^{b_1(a_1}_{(2,1,0)}
x^{a_2} g^{a_3a_4)} x^2 x_{b_1}
- \lambda^{b_1b_2}_{(2,1,0)} g^{(a_1a_2)} g^{a_3a_4)} x^2 x_{b_1} x_{b_2}+
\lambda^{(a_1}_{(3,0,0)} x^{a_2} x^{a_3} x^{a_4)}
\nonumber\\
\phantom{n=4, \quad}{}
-\lambda^{b_1}_{(3,0,0)} x^{(a_1} x^{a_2} g^{a_3a_4)}  x_{b_1}-
\lambda^{(a_1}_{(3,0,0)} x^{a_2} g^{a_3a_4} x^2
+ \lambda^{b_1}_{(3,0,0)} g^{(a_1a_2} g^{a_3a_4)} x^2 x_{b_1}
\nonumber\\
\phantom{n=4, \quad}{}
+\lambda_{(4,0,0)} x^{a_1} x^{a_2} x^{a_3} x^{a_4}
-2\lambda_{(4,0,0)} x^{(a_1} x^{a_2} g^{a_3a_4)} x^2+
\lambda_{(4,0,0)} g^{(a_1a_2} g^{a_3a_4)} x^4.
\end{gather}

Substituting (47)--(54) into (6), (7) and carrying differentiation
operators to the right, we obtain explicit form of the
corresponding symmetry operators. For $n=1$ we have a complete set
of symmetry operators of the following form:
\begin{gather}
 Q^a_1=P_a=i\frac{\p}{\p x^a}, \quad
Q^{ab}_{1}=J_{ab}=x_aP_b-x_bP_a.
\end{gather}

We do not adduce explicit form of symmetry operators for $n>1$
because of the corresponding formulae being extremely cumbersome
(in fact, these expressions are given by relations (6), (7),
(47)--(54)).

First order symmetry operators $Q_1$ adduced in (55), form a Lie
algebra $AP(p,q)$, satisfying the following commutation relations:
\begin{gather}
[P_a, P_b]=0, \quad [P_a, J_{bc}]=i(g_{ab}P_c-g_{ac}P_b), \nonumber\\
[J_{ab}, J_{cd}]=i(g_{ac} J_{bd}+g_{bd} J_{ac}-g_{ac} J_{bd}-g_{bd} J_{ac}).
\end{gather}

It is easy to notice using the representation (36) that symmetry
operators of arbitrary order are polynomials of the operators
(55). In other words, all symmetry operators of finite order of
equation (1) belong to the enveloping algebra of the algebra
$AP(p,q)$.

\section{Explicit form of Killing tensor of arbitrary rank $\boldsymbol{j}$}

Calculation of conformal Killing tensors of rank $j$ (that is
construction of the general solution of equation (14)) may be done
similarly to what was presented above in Sections 3--5.
Construction of such solution simplifies utilization of the result
formulated in the following lemma.

\medskip

\noindent
{\bf Lemma 2.} {\it Let $F^{a_1a_2\ldots a_{j_0}}$ be an arbitrary solution
of the system {\rm (14)} for $j=j_0$, and $F^a$ be a solution of this
system for $j=1$. Then the function
\begin{gather}
F^{a_1a_2\ldots a_{j_0+1}}=[F^{(a_1a_2\ldots a_{j_0}}F^{a_{j_0}+1)}]^{SL},
\end{gather}
where $[ \cdot ]^{SL}$ means the traceless part of the tensor in
the square brackets (see {\rm (18)}) is a solution of equations {\rm (14)} for
$j=j_0+1$.}

\medskip

{\bf Proof} can be done by a direct check.

We adduce below without proof the general solution of equations
(14)for $m \leq 4$ and arbitrary $j$.

By means of the reasoning similar to that in Section 3 we can show
that in the two-dimensional space equations (14) are reduced to
Cauchy--Riemann equations, and corresponding symmetry operators are
determined up to arbitrary analytical functions determining
independent components of a symmetric traceless tensor
$F^{a_1a_2\ldots a_j}$ (there are two such components for $j \ne
0$ and one for $j=0$ (that is for the case when then tensor
$F^{a_1a_2\ldots a_j}$ is reduced to a scalar).

For $m=3$ the number of independent solutions of equations (14) is
equal to
\begin{gather}
N^3_j=\frac{1}{3}(j+1)(2j+1)(2j+3).
\end{gather}

Solutions are numbered by the pair of integers $c=(c_1,c_2)$
satisfying the conditions
\begin{gather}
0 \leq c_1 \leq j, \quad 0 \leq c_2 \leq 2c_1,
\end{gather}
and for each $c_1$ contain  $(2c_1+1)$ arbitrary parameters giving
independent components of symmetric traceless tensor $\lambda^{a_1
a_2 \dots a_{c_1}}$ of rank $c_1$. Explicit form of the
corresponding solutions is given by the formula
\begin{gather}
F^{a_1a_2 \ldots a_j}_{(c_1c_2)}= \left[ \varepsilon_{c_2} f^{a_1a_2 \ldots a_j}_{(c_1c_2)}+
(1-\varepsilon_{c_2}) f^{b(a_1a_2 \ldots a_{j-1}}_{(c_1c_2)}
\varepsilon^{a_j)bc}x_c\right]^{SL},
\end{gather}
where
\begin{gather}
f^{a_1a_2 \ldots a_j}_{(c_1c_2)} =
\sum^{\{\frac{c_2}{2}\}}_{m=0}(-2)^m
\binom{\{\frac{c_2}{2}\}}{m} \lambda^{b_1b_2 \ldots
b_m(a_1a_2\ldots a_{c_1-m}}_{(c_1c_2)}
\nonumber\\
\phantom{f^{a_1a_2 \ldots a_j}_{(c_1c_2)} =}{}
\times  x^{a_{c_1-m+1}} x^{a_{c_1-m+2}} \ldots x^{a_j)} x_{b_1} x_{b_2} \ldots x_{b_m}
x^{2(\{\frac{c_2}{2}\}-m)},
\end{gather}
and the symbol $[ \cdot ]^{SL}$ means the traceless part of the
corresponding tensor; see (18) for $m=3$.

For $m=4$ the number of independent solutions of equations (14) is
equal to
\begin{gather}
N^4_j=\frac{1}{12}(j+1)^2(j+2)^2 (2j+3).
\end{gather}
The solutions are numbered by triples of integers
$c=(c_1,c_2,c_3)$ satisfying the conditions
\begin{gather}
 0 \leq c_1 \leq j, \quad -c_1 \leq c_2 \leq c_1, \quad 0 \leq c_3 \leq
\left\{\frac{c_1-|c_2|}{2} \right\},
\end{gather}
and for each $c$ contain $N_c$ arbitrary parameters where
\begin{gather}
 N_c= \left\{ \begin{array}{ll} (c_1+1)^2, &  c_1=|c_2|, \vspace{2mm}\\
2(|c_2|+2c_3+1)(2c_1-|c_2|-2c_3+1), & c_1 \not = |c_2|. \end{array} \right.
\end{gather}
These parameters determine independent components of an
irreducible tensor of rank $R=R_1+2R_2 $ where
\begin{gather}
R_1=|c_2|+2c_3, \quad R_2= c_1-|c_2|-2c_3,
\end{gather}
and explicit expressions for the corresponding solutions have the
form
\begin{gather}
F_c^{(a_1a_2\ldots a_j)}= \Bigg[ \sum^{m+c_3}_{i=0}(-1)^i
\binom{m+c_3}{i} (x^2)^i \nonumber\\
{}\times
\lambda^{b_1b_2\ldots b_{m-i+c_3}(a_1a_2 \ldots a_{|c_2|-m+i+c_3} [a_{|c_2|+i-m+1+c_3}d_1]
\ldots [a_{c_1-m+i-c_3}d_{c_1-|c_2|-2c_3}]} \nonumber\\
{}
 \times x^{a_{c_1-m+i-c_3+1}} x^{a_{c_1-m+i-c_3+2}}
 \cdots x^{a_j)} x_{b_1} x_{b_2}\cdots x_{b_{m-i+c_3}}\nonumber\\
{}\times x_{d_1} x_{d_2}\cdots x_{d_{c_1-|c_2|-2c_3}}\Bigg]^{SL}.
\end{gather}
Here $ \lambda^{b_1 \ldots b_{m-i+c_3a_1} \ldots
a_{|c_2|-m+i+c_3} [a_{|c_2|+i-m+1+c_3}{d_1}] \ldots
[a_{c_1-m+i-c_3} d_{c_1-|c_2|-2c_3}]}$ is an arbitrary ir\-re\-du\-cib\-le
tensor of rank  $R_1+2R_2$  ($R_1$ and  $R_2$ are given in (65)),
\begin{gather}
 m= \left\{ \begin{array}{rl} -c_2, & c_2 < 0, \vspace{1mm}\\
0, & c_2 \geq 0, \end{array} \right.
\end{gather}
$\binom{m+c_3}{i}$ is a binomial coefficient, and the symbol $[
\cdot ]^{SL}$ means the traceless part of the corresponding tensor;
see (18), (19) for $m=4$. Symmetrisation is implied over the
indices $a_1,\ldots, a_j$ in the righthand part (the sum over all
possible permutations).

Thus, we have found the explicit form of the conformal tensor of
rank $j$ for $m\leq 4$. The formula (66) determines the general
form  such tensor for arbitrary $m>3$, but at that the total
number of independent solutions of equations (14) cannot be
determined, in general, by the relation (62), but requires special
calculation for each value of $m$.

Let us point out that the general solution of equations (14) for
$m>2$ can be presented in the form
\begin{gather}
 F^{a_1a_2 \ldots a_j}= \Bigg[ \sum^j_{l,k=0} \sum^{j-l-k}_{i=0}
\lambda^{b_1b_2\ldots b_{j-l-k-i}(a_1a_2 \ldots a_{l+i}[a_{l+i+1}d_1] \ldots
[a_{l+i+k} d_k]} \nonumber\\
{}\times (-1)^i {\binom{j-l-k}{i}} (x^2)^i x^{a_{l+k+i+1}} \cdots
x^{a_j)} x_{d_1} x_{d_2} \ldots x_{d_k} x_{b_1} x_{b_2} \cdots
x_{b_{j-l-k-i}}\Bigg],
\end{gather}
where $ \lambda^{b_1b_2\ldots b_{j-l-k-i}a_1\ldots
a_{l+i}[a_{l+i+1}d_1] \ldots [a_{l+i+k} d_k]} $ is a tensor
symmetric with respect to permutation of the indices $b_1,  \ldots
, a_{l+j+k}$ and antisymmetric with respect to permutation of the
indices $a_{l+i+f}$ with $d_f, f=1,2, \ldots ,k$, with convolution
of this tensor by any three indices with an absolutely
antisymmetric vanishing. Decomposing such tensor into irreducible
tensors we come to formulae (58)--(67) giving solutions of
equations (14) for $m=3,4$.

\section{Examples of solutions and symmetry operators\\ for $\boldsymbol{n \leq 3}$}

Let us adduce an explicit form of the solutions obtained and
corresponding symmetry operators for $m \leq 4$ and $n \leq 3$.
Quantities of such solutions in accordance to (58) and~(62) are
adduced in Table 2.

\medskip

\centerline{\small {\bf Table 2.} Quantities of independent solutions of equations (14).}

\medskip

\centerline{\begin{tabular}{|c|ccc|}
\hline
$m \Big\backslash j$  &  $1$   &  $2$ &  $3$   \\
\hline
$3$    &  $10$   &  $ 35 $   &  $84$  \\
$4$    &  $15$   &  $84$   &  $300$ \\
\hline
\end{tabular}}

\medskip

Quantities of the corresponding symmetry operators of order $n$
can be obtained by summation of quantities of solutions from $j=0$
to $j=n$. We adduce the result in Table~3.

\medskip

\centerline{\small {\bf Table 3.} Quantities of order $n$ symmetry operators of equation
(1)}

\centerline{\small with $\varkappa=0$ in $m$-dimensional space.}

\medskip

\centerline{\begin{tabular}{|c|ccccc|}
\hline
$m \Big\backslash n$  &  $0$   &  $1$ &  $2$ & $3$  &  $4$  \\
\hline
$3$    &  $1$   &  $ 11$   &  $46$ &  $130$ &  $295$  \\
$4$    &  $1$   &  $16$   &  $100$ &  $400$  &  $1225$  \\
\hline
\end{tabular}}

\medskip

For $m=2$ the number of solutions of equations (14) (and the
number of the corresponding symmetry operators) is infinite as
they are determined up to arbitrary functions.

Explicit expressions for all independent solutions of equations
(14) for $m \leq 4$ and $n \leq 3$ are given by the following
formulae
$(F^{(j)}=F^{(a_1a_2 \ldots a_j)})$:

$m=2$
\begin{gather}
j=0, \quad F^{(0)}=\varphi^0(x_1, x_2);\nonumber\\
j > 0, \quad F^{11 \ldots 1}=(\varphi_j+\varphi^*_j)+i(\xi_j+\xi^*_j);\nonumber\\
\phantom{j > 0, \quad}{} F^{11 \ldots 1 2}=i(\varphi_j^*-\varphi_j)+\xi_j-\xi^*_j.
\end{gather}
Here $\varphi_j$ and $\xi_j$ are arbitrary analytical functions of
two variables $x_1$, $x_2$, and other components of the tensor
$F^{a_1 a_2  \ldots  a_j}$ are expressed through (69) using
properties of of zero trace and symmetry.

$m=3$
\begin{gather}
j=0,  \quad F^{(0)}=\lambda;\nonumber\\
j=1,  \quad  F^a_{(0,0)}=\lambda_{(0,0)}x^a; \quad F^a_{(1,0)}=\lambda^a_{(1,0)}; \quad
F^a_{(1,1)}=\varepsilon_{abc}\lambda^b_{(1,1)}x^c;\nonumber\\
\phantom{j=1,  \quad}{}
F^a_{(1,2)}=2\lambda^b_{(1,0)}x_b x^a-\lambda^a_{(1,2)}x^2;
\\
j=2, \quad F^{a_1a_2}_{(0,0)}=\lambda_{(0,0)}\left(x^{a_1}x^{a_2}
-\frac{1}{3} g^{a_1a_2}x^2\right);
\nonumber\\
\phantom{j=2, \quad}{}
F^{a_1a_2}_{(1,0)}=\lambda^{a_1}_{(1,0)}x^{a_2}+\lambda^{a_2}_{(1,0)}x^{a_1}-\frac{2}{3}
g^{a_1a_2}\lambda^b_{(1,0)}x^b;
\nonumber\\
\phantom{j=2, \quad}{}
F^{a_1a_2}_{(1,1)}=(x^{a_1}\varepsilon^{a_2}{}_{bc}
+x^{a_2}\varepsilon^{a_1}{}_{bc})x^b\lambda^c_{(1,1)};
\nonumber\\
\phantom{j=2, \quad}{}
F^{a_1a_2}_{(1,2)}=(x^{a_1}\lambda^{a_2}_{(1,2)}+x^{a_2}\lambda^{a_1}_{(1,2)})x^2-4x^{a_1}
x^{a_2}\lambda^b_{(1,2)}x_b+\frac{2}{3} g^{a_1a_2}\lambda^b_{(1,2)}x_b x^2;
\nonumber\\
\phantom{j=2, \quad}{}
F^{a_1a_2}_{(2,0)}=\lambda^{a_1a_2}_{(2,0)};
\nonumber\\
\phantom{j=2, \quad}{}
F^{a_1a_2}_{(2,1)}=(\varepsilon^{a_1bc}\lambda^{ba_2}_{(2,1)}+\varepsilon^{a_2 bc}
\lambda^{ba_1}_{(2,1)})x_c;
\nonumber\\
\phantom{j=2, \quad}{}
F^{a_1a_2}_{(2,2)}= \lambda^{a_1a_2}_{(2,2)}x^2-(x^{a_1}
\lambda^{a_2b}_{(2,2)}+x^{a_2} \lambda^{a_1b}_{(2,2)})x_b+ \frac{2}{3} g^{a_1a_2}
\lambda^{bc}_{(2,2)}x_bx_c;
\nonumber\\
\phantom{j=2, \quad}{}
F^{a_1a_2}_{(2,3)}=2(x^{a_1}\varepsilon^{a_2}{}_{bk}+
x^{a_2}\varepsilon^{a_1}{}_{bk})\lambda^{kd}_{(2,3)}x^b x_d-
(\varepsilon^{a_1}{}_{ck}\lambda^{a_2k}_{(2,3)}+\varepsilon^{a_2}
{}_{ck}\lambda^{a_1k}_{(2,3)})x^cx^2;
\nonumber\\
\phantom{j=2, \quad}{}
F^{a_1a_2}_{(2,4)}= \lambda^{a_1a_2}_{(2,4)}x^4-2(x^{a_1}
\lambda^{2c}_{(2,4)}+x^{a_2} \lambda^{a_1c}_{(2,4)}x_cx^2+4x^{a_1}x^{a_2}
\lambda^{ca}_{(2,4)}x_cx_d;
\\
j=3, \quad F^{a_1a_2a_3}_{(0,0)}= \lambda_{(0,0)}\left(x^{a_1} x^{a_2}x^{a_3}
-\frac{1}{10} g^{(a_1a_2}x^{a_3)}x^2\right);
\nonumber\\
\phantom{j=3, \quad}{}
F^{a_1a_2a_3}_{(1,0)}= \lambda^{(a_1}_{(1,0)}x^{a_2}x^{a_3)}-\frac{1}{5}(g^{(a_1a_2}
\lambda^{a_3)}_{(1,0)}x^2 + 2g^{(a_1a_2}x^{a_3)}\lambda^b_{(1,0)}x_b);
\nonumber\\
\phantom{j=3, \quad}{}
F^{a_1a_2a_3}_{(1,1)}= x^{(a_1}x^{a_2} \varepsilon^{a_3)}{}_{bc} x^b \lambda^{c}_{(1,1)}
-\frac{1}{5} g^{(a_1a_2}\varepsilon^{a_3)}{}_{bc} x^b \lambda^c_{(1,1)};
\nonumber\\
\phantom{j=3, \quad}{}
F^{a_1a_2a_3}_{(2,0)} = \lambda^{(a_1a_2}_{(2,0)} x^{a_3)}-\frac{2}{5}
g^{(a_1a_2} \lambda^{a_3)b}_{(2,0)} x_b;
\nonumber\\
\phantom{j=3, \quad}{}
F^{a_1a_2a_3}_{(2,1)} = x^{(a_1} \varepsilon^{a_2}{}_{bc}\lambda^{a_3)c}_{(2,1)}x_b-
\frac{1}{5} g^{(a_1a_2}\varepsilon^{a_3)}{}_{bc} \lambda^{cd}_{(2,1)}x^b x_d;
\nonumber\\
\phantom{j=3, \quad}{}
F^{a_1a_2a_3}_{(2,2)} = \lambda^{(a_1a_2}_{(2,2)} x^{a_3)}x^2-2x^{(a_1}x^{a_2}\lambda^{a_3)}_{(2,2)}
x_b+\frac{4}{5} g^{(a_1a_2}x^{a_3)}\lambda^{bc}_{(2,2)}x_bx_c;
\nonumber\\
\phantom{j=3, \quad}{}
F^{a_1a_2a_3}_{(2,3)} = \varepsilon^{(a_1}{}_{bc} x^{a_2}(2x^{a_3)} \lambda^{bd}_{(2,3)} x_d-
\lambda^{a_3)b}_{(2,3)}x^2)x^c;
\nonumber\\
\phantom{j=3, \quad}{}
F^{a_1a_2a_3}_{(2,4)} = x^{(a_1}(\lambda^{a_2a_3)}_{(2,4)}x^4
-4x^{a_2}\lambda^{a_3)b}_{(2,4)}x_b x^2
+4x^{a_2}x^{a_3)}\lambda^{kl}_{(2,4)}x_kx_l)
\nonumber\\
\phantom{j=3, \quad}{}
-\frac{2}{5}g^{(a_1a_2}(x^{a_3)}
\lambda^{kl}_{(2,4)}x_k x_l x^2- \lambda^{a_3)b}x_b x^4);
\nonumber\\
\phantom{j=3, \quad}{}
F^{a_1a_2a_3}_{(3,0)} = \lambda^{a_1a_2a_3}_{(3,0)};
\nonumber\\
\phantom{j=3, \quad}{}
F^{a_1a_2a_3}_{(3,1)} = \varepsilon^{(a_3}{}_{bc} \lambda^{a_1a_2)b}x_c;
\nonumber\\
\phantom{j=3, \quad}{}
F^{a_1a_2a_3}_{(3,2)} = \lambda^{(a_1a_2a_3)}_{(3,2)}x^2-2x^{(a_3}
\lambda^{a_1a_2)b}_{(3,2)} x_b+\frac{4}{5} g^{(a_1a_2} \lambda^{a_3)bc}_{(3,2)} x_b x_c;
\nonumber\\
\phantom{j=3, \quad}{}
F^{a_1a_2a_3}_{(3,3)} = \varepsilon^{(a_1}{}_{bc}\left(\lambda^{a_2a_3)b}_{(3,3)}x^2-2x^{a_2}
\lambda^{a_3)}_{(3,3)} x_b x_c
+\frac{2}{5} g^{a_2a_3)}\lambda^{bcd}_{(3,3)}x_b x_cx_d\right);
\nonumber\\
\phantom{j=3, \quad}{}
F^{a_1a_2a_3}_{(3,4)} = \lambda^{(a_1a_2a_3)}_{(3,4)}x^4-4x^{(a_1}
(\lambda^{a_2a_3)c}_{(3,4)}x_cx^2-x^{a_2}\lambda^{a_3)bc}_{(3,4)}x_b x_c)
\nonumber\\
\phantom{j=3, \quad}{}
-\frac{4}{5} g^{(a_1a_2}(2x^{a_3)} \lambda^{bcd}_{(3,4)}x_d-\lambda^{a_3)bc}_{(3,4)}x^2)x_b x_c;
\nonumber\\
\phantom{j=3, \quad}{}
F^{a_1a_2a_3}_{(3,5)} = \varepsilon^{(a_1}{}_{}(\lambda^{a_3a_2)b}_{(3,5)}x^4-4x^{a_2}
\lambda^{a_3)bd}_{(3,5)} x_d x^2+
4x^{a_2}x^{a_3)}\lambda^{bkl)b}_{(3,5)}x_kx_l)x^c;
\nonumber\\
\phantom{j=3, \quad}{}
F^{a_1a_2a_3}_{(3,6)} = \lambda^{(a_1a_2a_3)}_{(3,6)}x^6-6x^{(a_1}
\lambda^{a_2a_3)c}_{(3,6)}x_cx^4
\nonumber\\
\phantom{j=3, \quad}{}
+12 x^{(a_1} x^{a_2}\lambda^{a_3)bc}_{(3,6)}x_bx_cx^2-8x^{(a_1} x^{a_2}x^{a_3)}
\lambda^{bcd}_{(3,6)}x_bx_cx_d;
\end{gather}

$m=4$
\begin{gather}
j=0, \quad F^{(0)}= \lambda;\nonumber\\
j=1, \quad F^{(a)}_{(0,0,0)} = \lambda x^a;
\nonumber\\
\phantom{j=1, \quad}{}
 F^{(a)}_{(1,-1,0)} = \lambda^b x^a x_b-\lambda^a x^2;
\nonumber\\
\phantom{j=1, \quad}{}
F^{(a)}_{(1,0,0)} = \lambda^{[ad]} x_d;
\nonumber\\
\phantom{j=1, \quad}{}
F^{(a)}_{(1,1,0)} = \lambda^a;
\\
j=2, \quad F^{a_1a_2}_{(0,0,0)} = \lambda_{(0,0,0)}\left( x^{a_1} x^{a_2}-
\frac{1}{4} g^{a_1a_2}x^2\right);
\nonumber\\
\phantom{j=2, \quad}{}
F^{a_1a_2}_{(1,-1,0)} = 2 \lambda^b_{(1,-1,0)}x_b\left( x^{a_1} x^{a_2}-
\frac{1}{4} g^{a_1 a_2}x^2\right)-\lambda^{a_1}_{(1,-1,0)} x^{a_2}x^2
\nonumber\\
\phantom{j=2, \quad}{}
-\lambda^{a_2}_{(1,-1,0)}x^{a_1}x^2+\frac{1}{2}g^{a_1 a_2} \lambda^c_{(1,-1,0)}x_cx^2;
\nonumber\\
\phantom{j=2, \quad}{}
F^{a_1a_2}_{(1,0,0)} = \lambda^{[a_1d_1]}_{(1,0,0)} x^{a_2} x_{d_1}+
\lambda^{[a_2d_1]}_{(1,0,0)} x^{a_1} x_{d_1};
\nonumber\\
\phantom{j=2, \quad}{}
F^{a_1a_2}_{(1,1,0)}=-\frac{1}{2}g^{a_1 a_2} \lambda^c_{(1,1,0)}x_c+
\lambda^{a_1}_{(1,1,0)}x^{a_2}+ \lambda^{a_2}_{(1,1,0)}x^{a_1};
\nonumber\\
\phantom{j=2, \quad}{}
F^{a_1a_2}_{(2,-2,0)}=2 \lambda^{b_1b_2}_{(2,-2,0)} x^{a_1} x^{a_2} x_{b_1}x_{b_2}-2
\lambda^{b_1a_1}_{(2,-2,0)} x^{a_2} x^2 x_{b_1}
\nonumber\\
\phantom{j=2, \quad}{}
-2 \lambda^{b_1a_2}_{(2,-2,0)} x^{a_1} x^2 x_{b_1}+2 \lambda^{a_1a_2}_{(2,-2,0)} x^4+\frac{1}{2}
g^{a_1 a_2}\lambda^{b_1b_2}_{(2,-2,0)} x^2 x_{b_1}x_{b_2};
\nonumber\\
\phantom{j=2, \quad}{}
F^{a_1a_2}_{(2,-1,0)}= \lambda^{b_1[a_1d_1]}_{(2,-1,0)} x^{a_1} x_{b_1}x_{d_1}+
\lambda^{b_1[a_2d_1]}_{(2,-1,0)} x^{a_1} x_{b_1}x_{d_1}
\nonumber\\
\phantom{j=2, \quad}{}
-\lambda^{a_1[a_2d_1]}_{(2,-1,0)} x^2 x_{d_1}- \lambda^{a_2[a_1d_1]}_{(2,-1,0)} x^2 x_{d_1};
\nonumber\\
\phantom{j=2, \quad}{}
F^{a_1a_2}_{(2,0,1)}=  \lambda^{ba_1}_{(2,0,1)} x^{a_2} x_b+
\lambda^{ba_2}_{(2,0,1)} x^{a_1} x_b - 2 \lambda^{a_1a_2}_{(2,0,1)} x^2-
\frac{1}{2} g^{a_1 a_2}\lambda^{bc}_{(2,0,1)} x_bx_c;
\nonumber\\
\phantom{j=2, \quad}{}
F^{a_1a_2}_{(2,0,0)}=  \lambda^{[a_1d_1][a_2d_2]}_{(2,0,0)} x_{d_1} x_{d_2};
\nonumber\\
\phantom{j=2, \quad}{}
F^{a_1a_2}_{(2,1,0)}= ( \lambda^{a_1[a_2d]}_{(2,1,0)} + \lambda^{a_2[a_1d]}_{(2,1,0)})x_d;
\nonumber\\
\phantom{j=2, \quad}{}
F^{a_1a_2}_{(2,2,0)}= \lambda^{a_1a_2}_{(2,2,0)};
\\
j=3, \quad F^{a_1a_2a_3}_{(0,0,0)}= \lambda_{(0,0,0)} \left(x^{(a_1} x^{a_2} x^{a_3)}-\frac{1}{2}
g^{(a_1a_2} x^{a_3)} x^2\right);
\nonumber\\
\phantom{j=3, \quad}{}
F^{a_1a_2a_3}_{(1,-1,0)}= \lambda^b_{(1,-1,0)} \left(x^{(a_1} x^{a_2} x^{a_3)}-\frac{1}{2}
g^{(a_1a_2} x^{a_3)} x^2\right)x_b-
\lambda^{(a_1}_{(1,-1,0)} x^{a_2} x^{a_3)} x^2
\nonumber\\
\phantom{j=3, \quad}{}
+ \frac{1}{3} g^{(a_1a_2} x^{a_3)}
\lambda^b_{(1,-1,0)}x_b x^2 +
 \frac{1}{6} g^{(a_1a_2}\lambda^{a_3)}_{(1,-1,0)}x^4;
\nonumber\\
\phantom{j=3, \quad}{}
F^{a_1a_2a_3}_{(1,0,0)}= \lambda^{([a_1d]}_{(1,0,0)} x^{a_2} x^{a_3)} x_d-\frac{1}{6}
g^{(a_1a_2}\lambda^{[a_3d])}_{(1,0,0)} x^2 x_d;
\nonumber\\
\phantom{j=3, \quad}{}
F^{a_1a_2a_3}_{(1,1,0)}= \lambda^{(a_1}_{(1,1,0)} x^{a_2} x^{a_3)} - \frac{1}{3}
g^{(a_1a_2} x^{a_3)} \lambda^{b}_{(1,1,0)} x_b -
\frac{1}{6} g^{(a_1a_2} \lambda^{a_3)}_{(1,1,0)}x^2;
\nonumber\\
\phantom{j=3, \quad}{}
F^{a_1a_2a_3}_{(2,-2,0)}= \lambda^{b_1b_2}_{(2,-2,0)} \left(x^{(a_1} x^{a_2} x^{a_3)}-\frac{1}{2}
g^{(a_1a_2}x^{a_3)}x^2\right)x_{b_1} x_{b_2}
\nonumber\\
\phantom{j=3, \quad}{}
-2\lambda^{b(a_1}_{(2,-2,0)} x^{a_2}x^{a_3)} x_bx^2
+\frac{2}{3}
g^{(a_1a_2}x^{a_3)} \lambda^{bc}_{(2,-2,0)} x_bx_cx^2+
\frac{1}{3} g^{(a_1a_2} \lambda^{a_3)b}_{(2,-2,0)} x_bx^4
\nonumber\\
\phantom{j=3, \quad}{}
+\lambda^{(a_1a_2}_{(2,-2,0)} x^{a_3)} x^4+\frac{1}{3} g^{(a_1a_2} \lambda^{a_3)b}_{(2,-2,0)} x_b x^4;
\nonumber\\
\phantom{j=3, \quad}{}
F^{a_1a_2a_3}_{(2,-1,0)}= \lambda^{b([a_1d]}_{(2,-1,0)} x^{a_2} x^{a_3)} x_b x_d-
\lambda^{(a_1[a_2d]}_{(2,-1,0)} x^{a_3)} x_d x^2
\nonumber\\
\phantom{j=3, \quad}{}
-\frac{1}{6} g^{(a_1a_2} \lambda^{[a_3d])b}_{(2,-1,0)} x^2 x_b x_d+
\frac{1}{6} g^{(a_1a_2} \lambda^{[a_3d])b}_{(2,-1,0)} x^2 x_b x_d;
\nonumber\\
\phantom{j=3, \quad}{}
F^{a_1a_2a_3}_{(2,0,0)}= \lambda^{([a_1d_1][a_2d_2]}_{(2,0,0)} x^{a_3)} x_{d_1} x_{d_2};
\nonumber\\
\phantom{j=3, \quad}{}
F^{a_1a_2a_3}_{(2,0,1)}= \lambda^{b(a_1}_{(2,0,1)} x^{a_2} x^{a_3)} x_b
-\frac{1}{3} g^{(a_1a_2} x^{a_3)} \lambda^{bc}_{(2,0,1)} x_b x_c
\nonumber\\
\phantom{j=3, \quad}{}
+ \frac{1}{6} g^{(a_1a_2} \lambda^{a_3)b}_{(2,0,1)} x_b x^2- \lambda^{(a_1a_2}_{(2,0,1)}
 x^{a_3)} x^2;
\nonumber\\
\phantom{j=3, \quad}{}
F^{a_1a_2a_3}_{(2,1,0)}= \lambda^{(a_1[a_2d]}_{(2,1,0)} x^{a_3)} x_d-
\frac{1}{6} g^{(a_1a_2} \lambda^{[a_3d])b}_{(2,1,0)} x_b x_d;
\nonumber\\
\phantom{j=3, \quad}{}
F^{a_1a_2a_3}_{(2,2,0)}= \lambda^{(a_1a_2}_{(2,2,0)} x^{a_3)}-
\frac{1}{3} g^{(a_1a_2} \lambda^{a_3)b}_{(2,2,0)} x_b ;
\nonumber\\
\phantom{j=3, \quad}{}
F^{a_1a_2a_3}_{(3,-3,0)}= \lambda^{b_1b_2b_3}_{(3,-3,0)}
\left(x^{(a_1} x^{a_2} x^{a_3)}  x_{b_1} x_{b_2} x_{b_3} +
 \frac{1}{2} g^{(a_1a_2} x^{a_3)} x^2 x_{b_1} x_{b_2} x_{b_3}\right)
\nonumber\\
\phantom{j=3, \quad}{}
 - \lambda^{(a_1a_2a_3)}_{(3,-3,0)} x^6-
3 \lambda^{b_1b_2(a_1}_{(3,-3,0)} x^{a_2} x^{a_3)}
x^2 x_{b_1} x_{b_2} + 3 \lambda^{b_1(a_1a_2}_{(3,-3,0)} x^{a_3)} x^4 x_{b_1}
\nonumber\\
\phantom{j=3, \quad}{}
- \frac{1}{2} g^{(a_1a_2} \lambda^{a_3)b_1b_2}_{(3,-3,0)} x_{b_1} x_{b_2} x^4;
\nonumber\\
\phantom{j=3, \quad}{}
F^{a_1a_2a_3}_{(3,-2,0)}= \lambda^{b_1b_2([a_1d]}_{(3,-2,0)} x^{a_2} x^{a_3)}
x_{b_1} x_{b_2} x_d+
\frac{1}{6} g^{(a_1a_2} \lambda^{[a_3d])b_1b_2}_{(3,-2,0)} x^2 x_{b_1} x_{b_2} x_d
\nonumber\\
\phantom{j=3, \quad}{}
-2 \lambda^{b(a_1[a_2d]}_{(3,-2,0)} x^{a_3)} x_b x_d x^2
+ \lambda^{(a_1a_2[a_3 d])}_{(3,-2,0)} x_d x^4;
\nonumber\\
\phantom{j=3, \quad}{}
F^{a_1a_2a_3}_{(3,-1,0)}= \lambda^{b_1([a_1d_1][a_2 d_2]}_{(3,-1,0)} x^{a_3)}
x_b x_{d_1} x_{d_2} -
 \lambda^{(a_1[a_2d_1][a_3 d_2])}_{(3,-1,0)} x_{d_1} x_{d_2} x^2;
\nonumber\\
\phantom{j=3, \quad}{}
F^{a_1a_2a_3}_{(3,0,0)}
= \lambda^{[a_1d_1][a_2 d_2][a_3d_3]}_{(3,0,0)} x_{d_1} x_{d_2} x_{d_3};
\nonumber\\
\phantom{j=3, \quad}{}
F^{a_1a_2a_3}_{(3,0,1)} = (\lambda^{b(a_1[a_2 d]}_{(3,0,1)} x^{a_3)}-
\frac{1}{6} g^{(a_1a_2} \lambda^{[a_3d])bc}_{(3,0,1)} x_c) x_b x_d-
\lambda^{(a_1a_2[a_3d_1])}_{(3,0,1)} x_{d_1} x^2;
\nonumber\\
\phantom{j=3, \quad}{}
F^{a_1a_2a_3}_{(3,-1,1)} = \left(\lambda^{b_1b_2(a_1}_{(3,1,1)} x^{a_2} x^{a_3)}-
\frac{1}{3} g^{(a_1a_2} x^{a_3)} \lambda^{b_1b_2c}_{(3,-1,1)} x_c\right.
\nonumber\\
\left.\phantom{j=3, \quad}{}
+
\frac{1}{2} g^{(a_1a_2} \lambda^{a_3)b_1b_2}_{(3,-1,1)} x^2\right) x_{b_1} x_{b_2}
-2 \lambda^{b(a_1a_2}_{(3,-1,1)} x^{a_3)} x_b x^2 + \lambda^{(a_1a_2a_3)}_{(3,-1,1)} x^4;
\nonumber\\
\phantom{j=3, \quad}{}
F^{a_1a_2a_3}_{(3,1,0)} = \lambda^{(a_1[a_2d_1][a_3d_2]}_{(3,1,0)} x_{d_1} x_{d_2};
\nonumber\\
\phantom{j=3, \quad}{}
F^{a_1a_2a_3}_{(3,1,1)} = \lambda^{b(a_1a_2}_{(3,1,1)} x^{a_3)} x_b-
\frac{1}{3} g^{(a_1a_2} \lambda^{a_3)bc}_{(3,1,1)} x_b x_c
 - \lambda^{(a_1a_2a_3)}_{(3,1,1)} x^2;
\nonumber\\
\phantom{j=3, \quad}{}
F^{a_1a_2a_3}_{(3,2,0)} = \lambda^{(a_1a_2[a_3d])}_{(3,2,0)} x_d;
\nonumber\\
\phantom{j=3, \quad}{}
F^{a_1a_2a_3}_{(3,3,0)} = \lambda^{a_1a_2a_3}_{(3,3,0)}
\end{gather}

Here $\lambda_{( \ldots )}$, $\lambda_{( \ldots )}^a$, $\lambda_{(
\ldots )}^{a_1 a_2}$, $\lambda_{(\ldots)}^{a_1 a_2 a_3}$ are
arbitrary symmetric tensors with zero trace.

Formulae (69)--(75) give explicit form of all linearly independent
conformal killing tensors of rank $j \leq 3$ in spaces of
dimension $p+q =2,3,4$ (for $p+q =2$ $j$ is arbitrary). To obtain an
explicit form of the corresponding operators it is sufficient to
substitute (69)--(75)into (6), (7) that is write down $j$-multiple
anticommutators of $F^{a_1a_2 \ldots a_j}$ with $\frac{\partial}{\partial x_{a_1}},
\frac{\partial}{\partial x_{a_2}},
\ldots,\frac{\partial}{\partial x_{a_j}}$.

\section{Killing tensors of rank $\boldsymbol{j}$ and order $\boldsymbol{s}$}

Until now we considered solutions of equations (11), (14) that
define Killing tensors (and conformal Killing tensors) of
arbitrary rank $j$, but only of the first order. In this section
we obtain explicit form of Killing tensors of rank $j$ and of
arbitrary order $s$. Such tensors are determined as general
solutions of equations (16).

The system of equations (16) is overdetermined including $N^m_{js}$
equations for $\hat N^m_{js}$ unknown variables, where
\begin{gather}
 N^m_{js}=\binom{j+s+m-1}{m-1}, \quad \hat N^m_{js}=\binom{j+m-1}{m-1},
\quad m=p+q.
\end{gather}

In the same way as it was done above in Section 3, we consider the
set of differential consequences of the system under consideration
that are obtained by $k$-multiple differentiation of every term of
the equation by $\frac{\partial}{\partial
x_{a_1}},\frac{\partial}{\partial x_{a_2}},
\ldots,\frac{\partial}{\partial x_{a_k}}$. This set is a system of
linear homogeneous algebraic equations of the following form:
\begin{gather}
F^{(a_1a_2\ldots a_j, a_{j+1} a_{j+2}\ldots  a_{j+s})  b_1 b_2 \ldots b_k} = 0,
\end{gather}
where the following derivatives are unknown variables:
\begin{gather}
 F^{a_1a_2\ldots a_j,  a_{j+1} a_{j+2} \ldots  a_{j+s}  b_1 b_2 \ldots b_k} \equiv
 \p^{a_{j+1}} \p^{a_{j+2}} \cdots \p^{a_{j+s}} \p^{b_1} \p^{b_2}\cdots  \p^{b_k}
F^{a_1a_2\ldots a_j}.
\end{gather}

The quantities of unknown variables $N^k_{\mbox{\scriptsize u}}$ and of equations
$N^k_{\mbox{\scriptsize e}}$ are equal to
\begin{gather}
 N^k_{\mbox{\scriptsize u}} =  \binom{j+m-1}{m-1}\binom{k+s+m-1}{m-1}, \quad
N^k_{\mbox{\scriptsize e}} = \binom{j+s+m-1}{m-1} \binom{k+m-1}{m-1},\!\!
\end{gather}
so conditions (24) are also fulfilled.

It can be shown (see attachment) that the system (77) is
non-degenerate, so it follows from (24) that
\[
F^{a_1a_2\ldots a_j, a_{j+1} a_{j+2} \ldots a_{j+s}  b_1 b_2 \ldots b_j} \equiv 0.
\]
Whence we conclude that the Killing tensors of rank $j$ and order
$s$ are polynomials of order $j+s-1$. It follows from (77), (79)
that such polynomial contains $n^m_{js}$ arbitrary parameters,
where
\begin{gather}
 n^m_{js}= \sum^{s-1}_{i=0} \binom{j+m-1}{m-1}
\binom{j+m-1}{m-1} + \sum^{s-1}_{k=0} (N^k_{\mbox{\scriptsize u}}-N^k_{\mbox{\scriptsize e}})
\nonumber\\
\phantom{n^m_{js}}{}
=\frac{s}{m}
\binom{j+m-1}{m-1} \binom{j+s+m-1}{m-1}.
\end{gather}

Here the first sum gives the number of independent solutions that
have order by $x$ smaller than $s$. Such solutions can be written
in the form
\begin{gather}
 F_i^{a_1a_2 \ldots a_j}=\lambda^{a_1a_2\ldots
a_j}_{b_1b_2\ldots b_i} x^{b_1} x^{b_2} \cdots x^{b_i}, \quad i < s,
\end{gather}
where $\lambda^{a_1a_2\ldots a_j}_{b_1b_2\ldots b_i}$ are numeric
parameters with no limitations set by equations (77) (certainly
the symmetry with respect to permutations of indices
$
a_\lambda \leftrightarrow a_\lambda'$, $b_\mu \leftrightarrow
b_\mu'$, $\lambda,\lambda'=1,2, \ldots, j$, $\mu,\mu'=1,2,\ldots,i$.


In the case $s=1$ the formula (80) is reduced to (25). In
particular, for $m=2,3,4$ we obtain from (80) that
\begin{gather}
n^2_{js}=\frac{1}{2} s(j+1)(j+s+1), \nonumber\\
n^3_{js}=\frac{1}{12} s(j+1)(j+2)(j+s+1)(j+s+2), \nonumber\\
n^4_{js}=\frac{1}{3!4!} s(j+1)(j+2)(j+3)(j+s+1)(j+s+2)(j+s+3),
\end{gather}

Thus we have determined the number of linearly independent Killing
tensors of rank $j$ and order $s$. To compute these tensors in
explicit form we will use the following two lemmas.

\medskip

\noindent
{\bf Lemma 3.} {\it  Let $F_i^{a_1a_2 \ldots a_j}$ be Killing tensors of rank
$j$ and order $s$, and  $\varphi$ be a functions satisfying the
equation
\begin{gather}
\partial^\mu \partial^\nu \varphi=0, \quad \mu,\nu=1,2 \ldots m.
\end{gather}
Then the function
\begin{gather}
 \tilde F^{a_1a_2 \ldots a_j}=\varphi F^{a_1a_2\ldots a_j}
 \end{gather}
is a Killing tensor of rank $j$ and order $s$.}

\medskip

{\bf Proof} is reduced to direct check of the lemma statement that is
$(s+1)$-multiple differentiation of (84) by $\partial^{a_{j+1}}$,
$\partial^{a_{j+1}}$, \ldots, $\partial^{a_{j+s+1}}$ and
subsequent symmetrization of the obtained expression by $a_1$,
$a_2$, $a_{j+s+1}$ using relations (16), (83).

\medskip

\noindent
{\bf Lemma 4.} {\it Let $F^{a_1a_2 \ldots a_j}$ be a Killing tensors of rank
$j$ and order $s$. Then the convolution
\begin{gather}
 \tilde F^{a_1a_2 \ldots a_{j-1}} = F^{a_1a_2 \ldots
a_j}x_{a_j}
\end{gather}
is a Killing tensor of rank $j+1$ and order $s+1$.}

\medskip

{\bf Proof} is similar.

The adduced Lemmas provide an effective algorithm for construction
of Killing tensors of order $s$ from Killing tensors of order 1
found above in Section~4. The only difficulty in application of
this algorithm is the need to sort out all linearly independent
solutions of the system (16) (whose number is determined by the
formulae (80), (82), as, generally speaking, there are more
solutions of the form (84) than we need).

The general solution of equations (16) is determined in the
following theorem.

\medskip

\noindent
{\bf Theorem 3.} {\it Equations {\rm (16)} in the space of dimension $m \leq 4$
have $n^m_{js}$ linearly independent solutions where $n^m_{js}$ is
given by the formula {\rm (82)}. These solutions have the form
\begin{gather}
F^{a_1a_2 \ldots a_j}_{(s)} = g^{(a_{j-1}a_j} F^{a_1a_2 \ldots a_{j-2})}_{(s)}+
\varepsilon_j \hat f^{a_1a_2 \ldots a_j}\nonumber\\
\phantom{F^{a_1a_2 \ldots a_j}_{(s)} =}{}
+ \sum^s_{d=1}  x^{a_{j+1}} x^{a_{j+2}} \cdots
x^{a_{j+d-1})}, \quad \varepsilon_j =\frac{1}{2}[1+(-1)^j],
\end{gather}
where $F^{a_1a_2 \ldots a_{j+d-1}}$ are Killing tensors of
rank $j+d -1$ and of order 1 whose explicit form is given by
Theorem 2, $F^{a_1a_2 \ldots a_{j-2}}_{(s)}$ are Killing tensors
of rank $j-2$ and of order $s$;
\begin{gather}
\hat f^{a_1a_2 \ldots a_j} = \sum^{\frac{j}{2}-1}_{\mu=0}
(-1)^\mu \binom{\frac{j}{2}-1}{\mu} x^{(a_1}x^{a_2} \cdots
x^{a_{2\mu+1}}
\nonumber\\
\phantom{\hat f^{a_1a_2 \ldots a_j} =}{}
\times g^{ a_{2\mu+2} a_{2\mu+3}} \cdots g^{a_{j-2}a_{j-1}} \lambda^{[a_j),c]}x_c,
\end{gather}
$\lambda^{[a_j,c]}$ is an arbitrary antisymmetric tensor of rank $2$.}

\medskip

\noindent
{\bf Proof.} By virtue of Lemmas 3, 4 the function $F^{a_1a_2 \ldots
a_j}_{(s)}$ given by formula (86) is a Killing tensor of rank $j$
and order $s$; the first term --- by definition, the second --- in
accordance to Lemma 3 (being the product of of a Killing tensor of
order 1 and $\varphi=\lambda^\mu x_\mu$), the third~--- in
accordance to Lemma~4 (each convolution with $x_\mu$ lowers the
rank and increases the order of a Killing tensor, and we make the
first convolution with the first-order tensors described above).

It is to some extent more difficult to make sure that formula (86)
gives all linearly independent Killing tensors of order $s$. Proof
of linear independence of all terms of the formula (86) is reduced
to comparison of terms having the same order by $x_{a_i}$ using
different convolutions by one, two etc. pairs of indices.
Calculation of then umber of independent solutions given by
formula (86) can be done easily by sorting through independent
solutions for first-order tensors $F^{a_1a_2 \ldots a_{j+d
-1}}$ entering the last term (such tensors are described in
Theorem 2, but it is necessary to restrict consideration with
solutions corresponding to $c_2>d-1$, as others give zero
input into convolutions of (86)), and adding the number of
solutions of the form (87). The result obtained is in compliance
with the formula
\[
N=n^m_{js}-n^m_{j-2s},
\]
where $n^m_{js}$ is the total number of solutions given by the
formulae (82), $n^m_{j-2s}$ is the number of solutions of the form
$g^{(a_{j-1}a_j}F_s^{a_1a_2\ldots a_{j-2})}$ that is also given in
(82), $N$ is the total number of solutions under the summation
sign and of solutions of the form (87). We omit the corresponding
cumbersome calculations.

Formula (86) determines recurrent relations for calculation of
explicit form of a Killing tensor of rank $j$ and order $s$ from a
known tensor of order $s$ and rank $j-2$. Such calculations may be
easily checked starting from known killing tensors of order 1 and
arbitrary rank, see Theorem~2.

Let us adduce as an example explicit expressions for Killing
vectors of order $s \leq 3$ in three-dimensional space received
from general relations (86):
\begin{gather*}
s=1, \quad F^a_{(1)}=\lambda^a+\varepsilon^{abc}\eta_b x_c,
\\
s=2, \quad F^a_{(2)}= F^a_{(1)} + \lambda^{ab} x_b+\lambda x^a
+\varepsilon^{abc} \eta_{bd} x_c x^d+
\xi^ax^2-x^a\xi^b x_b,
\\
s=3, \quad F^a_{(3)}= F^a_{(2)} + \lambda^{abc} x_b x_c+\tilde \lambda^b x_b x^a+
\varepsilon^{abc} \eta_{bdl} x_c x^d x^l\\
\phantom{s=3, \quad}{} + \xi^{ab} x_b x^2-x^a\xi^{bc}x_bx_c+
\varepsilon^{abc} x_b \xi_c x^2.
\end{gather*}
Here $\varepsilon^{abc}$ is a unit antisymmetric tensor, and other
Greek letters designate arbitrary symmetric tensors with zero
trace.

\section{Conformal Killing tensors of rank $\boldsymbol{j}$ and order $\boldsymbol{s}$}

Let us briefly discuss equations (17) describing conformal Killing
tensors of rank $j$ and order $s$, and adduce without proof
solutions of these equations for arbitrary $j$, $s$ and $m \leq 4$.

A constructive way for finding solutions of equations (17) is shown
by the following statement that may be proven by direct check.

\medskip

\noindent
{\bf Lemma 5.} {\it Let $F^{a_1a_2 \ldots a_j}_s$ be a conformal Killing
tensor of rank $j$ and order $s$, and $\varphi$ be an arbitrary
function satisfying the equation
\begin{gather}
 \partial^\mu \partial^\nu \varphi=g^{\mu\nu}\lambda, \quad
\lambda= \mbox{\rm const}.
\end{gather}
Then the function
\begin{gather}
\tilde F^{a_1a_2 \ldots a_j}_{s+1} = \varphi F^{a_1a_2 \ldots a_j}_s
\end{gather}
is a conformal Killing tensor of rank $j$ and order $s+1$.}

\medskip

We can show that quantities of linearly independent solutions of
equations (17) for arbitrary $j$, $s$ and $m =3,4$ are given by
formulae
\begin{gather}
m=3, \quad \hat N^3_{js}=\frac{s}{6}(2j+1)(2j+2s+1)(2j+s+1),\nonumber\\
m=4, \quad \hat N^4_{js}=\frac{s}{12}(j+1)^2(j+s+1)^2(2j+2+s).
\end{gather}

Using Lemma 5 we managed to construct $N^m_{js}$ linearly
independent Killing tensors of rank $j$ and order $s$ (giving full
system of solutions of equations (17)) in the following form:
\begin{gather}
F^{a_1a_2 \ldots a_j}_s =  \sum^s_{i=1}\left(F_i^{a_1a_2 \ldots a_j}
(x^2)^{i-1}+ \sum^{s-i}_{d=0} f^{a_1a_2 \ldots a_j}_{i-1
d}(x^2)^d\right).
\end{gather}
Here $F_i^{a_1a_2 \ldots a_j}$ are conformal tensors of rank $j$
and order 1 given by formulae (60), (61) or (66) (the index ``$i$''
distinguishes independent solutions of (91) with various degrees
of~$x^2$),
$f^{a_1a_2 \ldots a_j}_{i-1 \, d}$ are tensors of rank $j$
whose explicit form is adduced below.

In the case $m=2$ the number of conformal Killing tensors of order
1 appears to be infinite, see (69). The same formulae (69) give a
general form of a conformal Killing tensor of arbitrary order $s$.

In the case $m=3$ functions $f^{a_1a_2 \ldots a_j}_{i-1 \ d}$
are characterized by an additional integer $c$,
\begin{gather}
 0 \leq c  \leq 2j,
 \end{gather}
and are determined up to arbitrary symmetric zero-trace tensor
$\tilde \lambda^{a_1 a_2 \ldots a_R}$ of rank $R=2(c+i)-1$.
Explicit form of these functions is given by by formula (93):
\begin{gather}
 f^{a_1a_2 \ldots a_j}_{id} = \left[\varepsilon_c \hat
f^{a_1a_2 \ldots a_j}_{id c}+ (1-\varepsilon c) \hat f^{b
(a_1a_2 \ldots a_{j-1}}\varepsilon^{aj)bc}x_c\right]^{SL},
\end{gather}
where
\begin{gather}
\hat f^{a_1a_2 \ldots a_j}_{id c} =
\sum^{\{\frac{c}{2}\}}_{n=0}(-2)^n \binom{\{\frac{c}{2}\}}{n}
\tilde \lambda^{b_1 b_2 \ldots b_{d+n}(a_1a_2 \ldots a_{j-n}}
\nonumber\\
\phantom{\hat f^{a_1a_2 \ldots a_j}_{id c} =}{}
\times x^{a_{j-n+1}} x^{a_{j-n+2}}\cdots x^{a_j)}  x_{b_1} x_{b_2} \cdots x_{b_{d+n}}
x^{2(\{\frac{c}{2}\}-n)}, \quad \varepsilon_c=\frac{1}{2}[1+(-1)^c]
\end{gather}
with the symbol $[ \cdot ]^{SL}$ designating the zero trace part of
the corresponding tensor, see (18), (19) for $m=3$, and the index
$d$ is introduced for numeration of linearly independent
solutions of (91) with different degrees of $x^2$.

In the case $m=4$ the functions $f^{a_1a_2 \ldots a_j}_{id
c}$ are characterized by a pair of additional indices
$c=(c_1,c_2)$
\begin{gather}
 -j \leq c_1 \leq j, \quad 0 \leq c_2 \leq \left\{ \frac{j-|c_1|}{2} \right\}
 \end{gather}
and is determined up to arbitrary irreducible tensor $\tilde
\lambda^{a_1 a_2 \ldots a_{R_1}[a_{R_1+1}b_1] \ldots
[a_{R_1+R_2}b_{R_2}]}$ of rank $R_1+2R_2$ where
\begin{gather}
R_1=|c_1|+2c_2+i, \quad  R_2=j-|c_1|-2c_2.
\end{gather}
Explicit form of these functions is given by the formula (97):
\begin{gather}
f^{a_1a_2 \ldots a_j}_{id}
=\Bigg[ \sum^{n+c_2}_{d=0}(-1)^d \binom{n+c_2}{d}
(x^2)^d  \lambda^{b_1b_2 \ldots b_{n+c_2-d-i}(a_1a_2 \ldots
a_{|c_1|-n+d+i+c_2+1}d_1] \ldots} \nonumber\\
{}{}^{\ldots [a_{j-n+i+d-c_2} d_{j-|c_1|-2c_2}]}
x^{a_{j-n+i+d-c_2+1}}\cdots x^{a_j} x_{b_1} \cdots x_{b_{n+c_2-d-i}} x_{d_1}
 \cdots x_{d_{j-|c_1|-2c_2}} \Bigg]^{SL},
\end{gather}
where
\begin{gather}
n= \left\{ \begin{array}{rl} -c_1,  & c_1 < 0,\vspace{1mm}\\
0, & c_1 \geq 0, \end{array} \right.
\end{gather}
and symmetrization is implied in the right-hand part by indices
$a_1, a_2, \ldots, a_j$.

The formulae (91)--(98) give in explicit form all linearly
independent conformal Killing tensors of rank $j$ and order $s$ in
space of dimension $m=p+q \leq 4$. In particular, conformal
vectors of order $s \leq 3$ in three-dimensional space, in
accordance to (91), (94) have the following form:
\begin{gather*}
s=1, \quad F^a_{(1)}=\lambda^a_{(1)}
+\varepsilon^{abc}\eta^b_{(1)} x_c+ \xi^a_{(1)} x^2-2x^a \xi^b_{(1)} x_b+
\mu x^a;
\\
s=2, \quad F^a_{(2)}=F^a_{(1)}+\tilde F^a_{(1)} x^2 +
\lambda^{ab}_{(2)} x_b+\varepsilon^{abc} \eta^{bd}_{(2)} x_c x_d+
\xi^{ab}_{(2)} x^2 x_b-2x^a \xi^{bc}_{(2)} x_b x_c;
\\
s=3, \quad F^a_{(3)}=F^a_{(2)}+x^4 {\mathop{F}\limits^\approx}{}^a_{(1)}+ x^2 (
\lambda^{ab}_{(3)} x_b+ \varepsilon^{abc} \eta^{bd}_{(3)} x_c x_d +
\xi^{ab}_{(3)}x_b x^2-2x^a \xi^{bc}_{(3)} x_b x_c) \\
\phantom{s=3, \quad}{}+ \lambda^{abc}_{(3)} x_b x_c+
 \varepsilon^{abc} \eta^{bdk}_{(3)} x_c x_d x_k + \varepsilon^{abc}_{(3)}
x_bx_cx^2-2x^a \xi^{bcd}_{(3)} x_b x_c x_d.
\end{gather*}
Here $\varepsilon^{abc}$ is the unit antisymmetric tensor, and other
Greek letters designate arbitrary symmetric traceless tensors
$F^a_{(1)}$, $\tilde F^a_{(1)}$ and ${\mathop{F}\limits^\approx}{}^a_{(1)}$
are first-order Killing vectors (generally speaking, different).

\section{Conclusion}

Let us sum up. We have defined the notion of Killing
tensor of rank $j$ and order $s$ and of conformal Killing tensor
of rank $j$ and order $s$. These tensors are defined as general
solutions of equations (16) or (17) that in the case $s=1$
coincide with generally accepted equations for Killing tensors and
conformal Killing tensors, see e.g.~\cite{Walker}.

We limit ourselves with investigation of equations (16) and (17)
in flat de Sitter space, and generalization of these equations for
for the case of spaces with non-zero curvature requires
replacement of $\partial^{a_i}$ for covariant derivatives.

Equations (16) and (17) are natural generalizations of the Killing
equations \cite{Ibragimov, Killing} and arise in
description of higher-order symmetry operators. In the present
paper we show relation of these equations (for first-order
tensors) with higher-order symmetry operators of Klein--Gordon--Fock
equation, see Sections 2, 5. Equations for Killing tensors and
conformal Killing tensors of order $s < 1$ arise in problems of
description of symmetry operators of order $s$ for systems of
partial differential equations --- in particular, for the Maxwell
equations \cite{F4}. We have found in explicit form all
non-equivalent Killing tensors of rank $j$ and order $s$ in the
space of dimension $p+q$ for arbitrary $j$ and $s$ and $p+q \leq
4$. Limitation by dimension of space is based, on the one hand, on
practical reasons (the absolute majority of equations of
mathematical physics being the field of research interests of the
authors, have dimension $m \leq 4$ with respect to independent
variables), and, on the other side, on difficulties that had not
been overcome to the moment in proof of non-degeneracy of systems
of algebraic equations for coefficients of Killing tensors in
spaces of arbitrary dimension, see attachment. At that formulae
(42), (66), (86) giving solutions of equations (16), (17) for
$p+q=4$, probably give the general solution of these equations for
arbitrary $p+q \geq 4$.

The found general solutions for Killing tensors and conformal
Killing tensors of arbitrary order and rank may find quite large
use in description of symmetry operators of systems of partial
differential equations. In this paper using these solutions we
found full set of symmetry operators of arbitrary finite order for
Klein--Gordon--Fock equations with zero and non-zero mass.

\appendix

\section{Non-degeneracy of systems\\ of equations for
coefficients of Killing tensors}

\renewcommand{\theequation}{A.\arabic{equation}}
\setcounter{equation}{0}

We will adduce proof of Theorem 1 stating non-degeneracy of system
of linear algebraic equations (22). As it is cumbersome we adduce
it in abridged form.

The main difficulty of the analysis of system (22)
is the need to do it for arbitrary value of $j$, that is for a
system of arbitrary fixed dimension given by formula (20).

Let us consider equations (22) for $m=4$, at that equations for
$m<4$ will be included into the analysis as particular cases.
Indices $a_1, a_2, \dots, a_{j+1}$ and $b_1, b_2, \dots,
b_{k}$ with $k \leq j$ independently take values from 1 to 4. At
that, as it is easy to notice, the system (22) splits at
non-linked subsystems $M(s_1,s_2,s_3,s_4)$, where $s_l$
$(l=1,2,3,4)$ gives the number of indices having the value $l$. It
is obvious that
\begin{gather}
 s_1+s_2+s_3 + s_4 = j+1+k,
 \end{gather}
so $0 \leq s_l \leq j+k+1$.

System (22) is non-degeneracy iff all its subsystems $M(s_1,s_2,s_3,s_4)$ are
non-degeneracy. Without loss of generality, for arbitrary subsystem $M(s_1,s_2,s_3,s_4)$
we can put
\begin{gather}
s_1 \leq s_2 \leq s_3 \leq s_4,
\end{gather}
other cases can be reduced to (A.2) by renumeration of variables.

Let us prove non-degeneracy of an arbitrary subsystem $M(s_1, s_2,
s_3,s_4)$.

We designate by the symbol $n_l$ $(l=1,2,3,4)$ the number of indices
of unknown variable $F^{a_1 a_2 \ldots a_j, a_{j+1} b_1 b_2 \ldots
b_k}$ present on the left of the comma and equal to $l$, and by
the symbol $m_l$~--- the number of indices after the comma that are
equal to $l$. Obviously, the following should be satisfied,
\begin{gather}
 m_c+n_c=s_c, \quad n_1+n_2+n_3+n_4=j, \quad m_1+m_2+m_3+m_4=k+1,
 \end{gather}
so out of eight numbers $n_l$ and $m_l$ only three will be
linearly independent (see (A.1)). Let us choose the following
numbers as independent: $n_1$, $n_2$ and $n_3$, then the triple
$(n_1,n_2,n_3)$ will completely determine a vector $F^{a_1 a_2 \ldots
a_j, a_{j+1} b_1 b_2 \ldots b_k}$ from the subsystem
$M(s_1,s_2,s_3,s_4)$. Using for such vector the designation
$F(n_1,n_2,n_3)$ and considering relations (A.1)--(A.3), we can write
any equation (22) from the subsystem $M(s_1,s_2,s_3,s_4)$ in one
of the following forms:
\begin{gather}
(n_3+1)F(0,0,n_3)+(j-n_3)F(0,0,n_3+1)=0, \nonumber\\
\max\{s_3-k-1,-1\} \leq n_3 \leq s_3-1;\\
(n_2+1)F(0,n_2,n_3)+n_3F(0,n_2+1, n_3-1)+(j-n_2-n_3)F(0,n_2+1,n_3)=0, \nonumber\\
\max\{0, s_1+s_2-k-1\} \leq n_2 \leq s_2-1, \nonumber\\
\max\{0, s_1+s_2+s_3-k-1-n_2 \} \leq n_3 \leq \min\{s_3, j-n_2\};
\\
(n_1+1)F(n_1,n_2,n_3)+n_2F(n_1+1,n_2-1,n_3)+n_3F(n_1+1,n_2,n_3-1)\nonumber\\
\qquad {}+(j-n_1-n_2-n_3) F(n_1+1, n_2, n_3)=0,  \nonumber\\
\max \{0, s_1-k-1\} \leq n_1 \leq s_1-1, \quad
\max \{0, s_1+s_2-k-1-n_1\} \leq n_2 \leq s_2, \nonumber\\
\max \{0, s_1+s_2+s_3-k-1-n_1-n_2\} \leq n_3 \leq
\min\{s_3, j-n_1-n_2\}.
\end{gather}

When $s_1>0$ (the case $s_1=0$ is considered below) there are
three possibilities:
\begin{gather}
1. \quad s_1+s_2 < k+1,\nonumber\\
2. \quad s_1+s_2 \geq k+1, \quad s_1 <k+1, \nonumber\\
3. \quad s_1 \geq k+1.
\end{gather}

Let us consider these possibilities one by one.

In the case 1 the system under consideration is given by the
formulae (A.4)--(A.6).

Let us present the vector $F(n_1, n_2, n_3)$ in the form of a
column whose components are numbered by the index
\begin{gather}
 F(n_1, n_2, n_3)= \big(F(n_1, n_2, \tilde n_3), F(n_1, n_2, \tilde n_3+1) \ldots
F(n_1, n_2, \hat n_3)\big)^T,
\end{gather}
where $\tilde{n_3}$ and $\hat{n_3}$ are minimal and maximal values
of $n_3$, and each vector $F(n_1, n_2, \tilde n_3 +k)$, in its
turn, will be regarded as a column whose components are numbered
by the index~$n_2$, $0 \leq n_2 \leq s_2$. Then it is possible to
write equations (A.4)--(A.6) in the matrix form:
\begin{gather}
AF=0,
\end{gather}
where
\begin{gather}
 A=\left( \begin{array}{ccccc}
B_0  &        &       &  & \\
E_1  &  B_1   &       &  & \\
     &  2E_2  & B_2   &  & \\
   &  & \ddots & \ddots  &  \\
     &         &       & s_1E & B_{s_1}
\end{array} \right),\nonumber\\
B_l=\left( \begin{array}{ccccc}
D_l  &        &       &  & \\
E_{l1}  &  D_{l+1}   &       &  & \\
&  2E_{l2}  & D_{l+2}   &  & \\
   &  & \ddots & \ddots  &  \\
     &        &  & s_2E_{ls_2} & D_{l+s_2}\end{array} \right).
\end{gather}

Here $E_k$ and $ E_{lf} $ ($l, k=1,2, \ldots, s_1$, $f=1,2, \ldots, s_2)$
are unit matrices whose number of rows coincides with the number
of rows of the adjacent matrices on the right (the number of
columns of $B_k$ ($D_l$) coincides with the number of rows of
$B_{k+1}$ ($D_{l+1}$)), $D_l$ being the matrices whose explicit
form is determined below. Namely, with $s_1+s_2+s_3 <k+1$:
\begin{gather} D_l=\left(\begin{array}{ccccc}
j-l+1  &        &       &  & \\
1  &  j-l  &       &  & \\
     &  2  & j-l-1   &  & \\
   &  & \ddots\quad & \ddots  &  \\
     &         &       & s_3 & j-l+1-s_3
\end{array} \right)
\end{gather}
with $s_1+s_2+s_3 \geq k+1$
\begin{gather} D_l=\left(\begin{array}{ccccc}
a_1  & j-a_1-l+1  &       &  & \\
  &  a_1+1  & j-a_1-l      &  & \\
   &  & \ddots\qquad  & \ddots  &  \\
     &         &       & a_2 & j-a_2-l+1
\end{array} \right),
\end{gather}
where
$a_1 = \max\{0, s_1+s_2+s_3-k-l\}$, $ a_2 = \min\{s_3, j-l+1 \}$.

At that in the cases $a_1=0$ or (and) $s_3 \geq j-l+1$ in the
matrix (A.12) the first or (and) last column should be crossed out.

\medskip

\noindent
{\bf Note A.1.} It can be shown that there are always will be less of
matrices $D_l$ of the form (A.11) (or (A.12) $a_1\not=0$ $j-a_1-l+1 \neq
0$) than of the matrices (A.12) with $a_1=j-a_2-l+1\equiv 0$.

\medskip

Our task is to prove that all rows of the matrix $A$ (A.10) are
linearly independent.

Writing this matrix in the equivalent form
\[ A' = \left( \begin{array}{ccccc}
E_1  &  B_1      &       &  & \\
 &  2F_2   & B_2       &  & \\
 &  & \ddots & \ddots  &  \\
     &         &       & s_1E_{s_1} & B_{s_1} \\
B_0  &        &  &
\end{array} \right)
\]
and subjecting $A^\prime$ to the transformation $A^\prime
\rightarrow A^{\prime \prime}=V A^\prime W$ that does not change
the rank, with $V$ and $W$ being reversible matrices of the form
\begin{gather*}
V = \left( \begin{array}{ccccc}
E_1  &     &    &  & \\
 &  E_2    &    &  & \\
 &  & \ddots &   &  \\
 & & & E_{s_1} &  \\
-B_0  & B_0B_1 & \ldots & (-1)^{s_1+1}\frac{1}{s_1!} B_0 B_1 B_2 \cdots B_{s_1} & E_0
\end{array} \right),\\
 W = \left(  \begin{array}{ccccccc}
E_1&-B_1& \frac{1}{2!}B_1B_2&-\frac{1}{3!}B_1B_2B_3&
 \ldots&(-1)^{s_1}\frac{1}{(s_1-1)!}B_1B_2 \ldots B_{s_1-1} \\
& E_2 & -B_2 & \frac{1}{2!}B_2B_3&\ldots &(-1)^{s_1-1}\frac{1}{(s_1-2)!}B_2B_3
\ldots B_{s_1-1} \\
 &  & \ldots & \ldots  & \ldots & \ldots  \\
 & & & & E_{s_1} & -B_{s_1-1}  \\
 &  &  & & & E_{s_1}
\end{array} \right),
\end{gather*}
we get
\begin{gather}
 A'' = \left( \begin{array}{ccccc}
E_1  &     &    &  & \\
 &  E_2    &    &  & \\
 &  & \ddots &   &  \\
 & & & E_{s_1} &  \\
 & & & & \frac{1}{s_1!} B_0 B_1 B_2 \cdots B_{s_1}
\end{array}
\right)
\end{gather}
and proof of linear independence of the rows of the matrix (A.10)
is reduced to proof of linear independence of the rows of the
matrix
\begin{gather}
\hat B^{s_1} = B_0 B_1 \cdots B_{s_1}.
\end{gather}

\noindent
{\bf Lemma A.1.} {\it The matrix $(A.14)$ can be split into blocks $\hat
B^{s_1}_{lf}$ including the same number of rows as the matrix
$D_{s_1+l-1}$, $1 \leq l \leq s_{2+1}$, $1 \leq f \leq s_{2+1}$.
These blocks have the following form:
\begin{gather}
 B^{s_1}_{lf} = \left\{ \begin{array}{l}
\hat 0, \ \ -s_2 \leq l-f < 0 \  \ \mbox{or} \ \ s_1+2 \leq l-f \leq s_2,  \vspace{2mm}\\
\displaystyle
\binom{s_{1}+1}{l-f} P^{l-1}_{l-f} D_{l-1} D_l \ldots D_{s_1+f-1}, \ \ 0 \leq l-f \leq s_1,
\vspace{2mm}\\
\displaystyle \binom{s_{1}+1}{s_{1}+1} P^{l-1}_{s_{1}+1} E, \  \ l-f=s_1+1. \end{array} \right.
\end{gather}
Here $D_{l-1}, D_l, \ldots$ are matrices from $(A.10)$, the general
form of which is given by the formulae {\rm (A.11)}, {\rm (A.12)},
$\hat 0$ and $E_k$ are zero and unity matrices of corresponding dimensions,
$\binom{m}{k}=\frac{m!}{k!(m-k)!}$, $P^k_m=\frac{k!}{(k-m)!}$.}

\medskip

{\bf Proof} can be done by induction, by successive investigation of the
products $B_0B_1$, $B_0B_1B_2, \ldots$.

\medskip

\noindent
{\bf Lemma A.2.} {\it  The matrix {\rm (A.14)} by finite number of elementary
transformations can be reduced to the form}
\begin{gather}
 \hat B^{s_1}=\left( \begin{array}{@{}c@{}c@{}c@{}c@{$\!\!\!\!$}c@{$\!\!\!\!$}c@{$\!\!\!\!$}c@{$\!\!\!\!$}c@{}}
 &    & &    &  & & & D_0D_1\ldots D_{s_1+s_2} \\
 &    & &     & &   & \cdots & \\
 &    & &     &   & D_{s_1-1}D_{s_1}\ldots D_{s_2+1} & & \\
&     & &    & D_{s_1}D_{s_1+1}\cdots D_{s_2} &  & & \\
 &     &&   E_{s_2-s_1}  &  &   &  & \\
 &  & \cdots  & &   &  & \\
 & E_2 & & &  &  &  \\
 E_1  &   & & & &  & &
\end{array}
\right)
\end{gather}

\noindent
{\bf Proof.} We can describe elementary transformations mentioned in the
lemma in the following way:

1. Let us go from the matrix (A.14) to the matrix $\Pi_d$ (with $d$
successively taking values $1,2,\ldots,s_2-s_1$), using the
following algorithm:

1) Represent the matrix (A.14) in the block form (A.15) and operate
with ``rows'' with the number $l$, including blocks
$\hat B^{s_1}_{l1},\hat B^{s_1}_{l2},\ldots,\hat B^{s_1}_{l \,
s_2+1}$.

2) Multiply $(t+1)$-th ``row''  of the matrix $\hat B^{s_1}$ by the
matrix $D_{t-1}$, and the ``row'' of the matrix $\hat B^{s_1}$ with
the number $t$ --- by the number $s_1-t+2$ and subtract the latter
from the former. Write the obtained result instead of the ``row''
with the number $t$. Perform this operation successively with all
``rows'' for $t=1,2,\ldots, s_1$.

3) Multiply $(s_1+2)$-th ``row'' of the matrix $\hat B^{s_1}$ by
$D_{s_1} \ldots D_{s_2} \ldots D_{s_1+d-1}$ and subtract from
the obtained result the ``row'' with the number $s_1+1$ multiplied
by $d$. Write the obtained result instead of the ``row'' with the
number $s_1$, leaving the remaining ``rows'' unchanged.

4) As a result of the described transformations $\hat B^{s_1}
\rightarrow \Pi^\prime_d$, where $\Pi^\prime_d$ is the matrix
having in the first column the only non-vanishing element
$(\Pi'_d)_{s_1+2 \, 1}$. It is possible to get by means of
elementary transformations that all elements of the $s_1+2$-th
``rows'' would also vanish (except $(\Pi'_d)_{s_1+2 \, 1}$).

5) Let us put the $s_1+2$-th ``row'' to the lowest position. As a
result we get the matrix $\Pi_{s_1-s_2}$ of the following form:
\begin{gather}  \Pi_{s_2-s_1} = \left( \begin{array}{ccccc}
 &     &    &  & M \\
 &     &    & E_{s_2-s_1} & \\
 &  & \cdots &   &  \\
 & E_2  & &  &  \\
E_1  & & & &
\end{array} \right),
\end{gather}
where $M$ is a matrix that can be split into blocks of the
following form:
\[
M_{lf}= \left\{ \begin{array}{l} 0, \ \  -s_1 \leq l-f < 0,\vspace{1mm}\\
\displaystyle \binom{s_{2}+1}{l-f} P^{l-f}_{l-1} D_{l-1} D_l \cdots D_{f+s_2+1}, \ \
0 \leq l-f \leq s_1, \end{array} \right.
\quad (f, l)=1,2, \ldots, s_{1+1}.
\]

2. Let us simplify the matrix $M$ by using successively the
adduced algorithm for $q=0,1, \ldots, s_1-1$, and for each value
of $q$ --- for $k=1,2, \ldots, s_1-q$.

1) Let us multiply the $(k+1)$-th ``row'' of the matrix $M$ by the
matrix $D_{k-1}$, and the $k$-th column --- by the number $s_1-k+2$,
and subtract the former from the latter, leaving other ``rows''
unchanged.

2) Perform this operation successively for all $k=1,2, \ldots,
s_1-q$, simplifying the matrices obtained at each step by means of
elementary transformations vanishing all elements of ``rows'' except
one that was the only non-vanishing in its column.

3) Perform operations 1), 2) successively for all $q=0,1, \ldots,
s_1-1$.

As a result we come to the matrix (A.16). The lemma is proved.

\medskip

\noindent
{\bf Lemma A.3.} {\it Let the matrix $ D=\|d_{ab}\|$, $a=1,2, \ldots, s$,
 $b=1,2 \ldots, r$, $s < r $ have the rang~$s$, with all minors $D$
being positive, and the matrix $B$ have the form
\begin{gather}  B = \left( \begin{array}{ccccc}
b_{11} &     &    &  &   \\
b_{21} & b_{22}    &    &  & \\
 & b_{32} & b_{33} & & \\
 &  & \ddots  & \ddots  &  \\
 & & & b_{r-1r-2} & b_{r-1r-1} \\
  &  & & & b_{rr-1}
\end{array}\right),
\end{gather}
where $ b_{kk} > 0 $  and $ b_{k+1\, k} >0$,  $k=1,2, \ldots, r-1 $.
Then the matrix $C=DB$ also has the rank $s$, and all minors $C$
are positive.}

\medskip

{\bf Proof} is reduced to direct utilization of the Binet--Cauchy formula
\cite{Lankaster} representing spinors of the matrix product $DB$
via sum of the products of minors of the matrix $D$ by minors of
the matrix $B$. As a result each minor of the matrix $C$ can be
represented as the sum of positive values. The Lemma is proved.

By virtue of the Lemma A.2 proof of linear independence of the
matrix $A$ (A.10) is reduced to proof of linear independence of
rows of the matrix ${\mathcal D}^d$,
\begin{gather}
{\mathcal  D}^d \equiv D_{s_1-d} D_{s_1-d+1} \cdots D_{s_2+d}, \quad
d=0,1,2, \ldots, s_1,
\end{gather}
where $D_{s_{1-d}}, D_{s_1-d+1}, \ldots$ are matrices of the form
(A.11) or (A.12). By virtue of Note~A.1 the number of
rows of each matrix (A.19) does not exceed the number of its
columns. Considering successively the products $D_{s_1-d}
D_{s_1-d+1}, D_{s_1-d}D_{s_1-d+1}D_{s_1-d+2},\ldots$ and using
each time either the Silvester inequality \cite{Korn} or Lemma
A.3, it is not difficult to show that the rank of the matrix (A.19)
coincides with the number of its rows, and, whence, all rows of
the matrix $A$ (A.10) are linearly independent.

We have proved non-degeneracy of of the system (A.4)--(A.6) for the
case 1 from (A.7). In the case 2 when $s_1+s_2 \geq k+1$, the
system (A.4)--(A.6) is reduced to equations (A.5), (A.6) that can also
be written in matrix form (A.9) where $A$ is given by (A.10), but
the blocks $B_l$ have a new form. Namely, with $s_2 <k+1$
\begin{gather}
  B_l = \left( \begin{array}{ccccc}
(s_1+s_2-k-l)E_0 & D_0    &    &  &   \\
 &  (s_1+s_2-k-l+1) E_1  & D_1 &  & \\
 &  & \ddots  & \ddots  &  \\
 & & & s_2E_{k-s_1+l} & D_{k-s_1+l} \end{array} \right),
\end{gather}
$l=0,1, \ldots, s_1+s_2-k-1$,
and
\begin{gather}  B_l = \left(\begin{array}{cccc}
D_{l+k-s_1-s_2} &  &  &   \\
 E_1 & D_{l+k-s_1-s_2+1}  &  &  \\
   &  \ddots  & \ddots  & \\
  & & s_2E_{s_2} & D_{l+k-s_1} \end{array} \right),
  \end{gather}
if $s_1+s_2-k \leq l \leq s_1$. If $s_2 \geq k$, then all matrices
$B_l$, $0 \leq l \leq s_1$, are given by the formula (A.20).

Explicit form for the matrices $D_l$ is given by relations
(A.22)--(A.23):

for $s_2<k$
\begin{gather}  D_l = \left(\begin{array}{ccccc}
a_1 & j-l-a_1  & & &   \\
 & a_{1+1}  & j-l-a_1-1 &&  \\
   & &  \ddots  & \ddots& \\
   & & & a_2  & j-l-a_2\end{array} \right),
 \end{gather}
where
$a_1=\max\{0,s_1+s_2+s_3-k-l-1\}$, $a_2=\min\{s_3, j-l\}$,
$l=0,1, \ldots k-d_1$,
\begin{gather}  D_l = \left(\begin{array}{@{}c@{\,\,}c@{$\!\!\!\!\!\!$}c@{$\!\!\!\!\!\!$}c@{}}
a_1 & j-l+k+1-s_1-s_2-a_1  &  &   \\
 & a_1+1  & j-l+k-s_1-s_2-a_1 &  \\
   &  &\ddots\qquad  &  \ddots\qquad   \\
   &  & a_2  & j-l+k-s_1-s_2-a_2\end{array}\right),\!\!
 \end{gather}
where
$a_1 = \max\{0,s_3-l \}$, $a_2=\min\{s_3, j-l+k+1-s_1-s_2 \}$,
$l=k-s_1+1, k-s_1+2,\ldots, k$.

If $a_1=0$ then first columns in (A.22) and (A.23) should be crossed
out, if $s_3 \geq j-l$, then the last column in (A.22) should be
crossed out, and with $s_3 \geq j-l+k+1-s_1-s_2$ it is necessary
to cross out the last column in (A.23).

In the case $s_2 \geq k$ the matrices $D_l$ are given by the
formula (A.22) for all $l$.

Our task is to prove linear independence of rows of the matrix $A$
determined by relations (A.10), (A.20)--(A.23). Transforming this
matrix to the form (A.13) we reduce this problem again to
investigation of the matrix (A.14) that in our case can be split
into blocks of the form
\begin{gather} \hat B^{S_1}_{lf} = \left\{ \begin{array}{l}
\hat 0, \ \ -s_2 \leq l-f < s_1+s_2-k \ \ \mbox{or} \ \
 k-s_2+2 \leq l-f \leq s_2,  \vspace{1mm}\\
\displaystyle \binom{s_1+1}{l+s_1+s_2-k-f} P^{l+s_1+s_2-k-1}_{l+s_1+s_2-k-f} D_{l-1} D_l \cdots
D_{k-s_2-1-f},\vspace{1mm}\\
\qquad -s_1-s_2+k \leq l-f \leq k-s_2,  \vspace{1mm}\\
\displaystyle \binom{s_1+1}{s_1+1} P^{l+s_1+s_2-k-1}_{s_1+1} E, \  \ l-f = k+1-s_2,
 \end{array} \right.
 \end{gather}
 where $1 \leq l \leq k+1-s_2$, $1 \leq f \leq s_2+1$.

Further proof is done in full analogy with the proof for the case~1.

Let us consider now the third case from (A.7). The corresponding
system of equations (A.4)--(A.6) is reduced to equations (A.6).
Writing these equations in the matrix form (A.9) we come to the
corresponding matrix $A$ of the following form:
\begin{gather}  A = \left(\begin{array}{ccccc}
(s_1-k)E_0 & B_0  &  &  & \\
 & (s_1-k+1)E_1  & B_1 & &  \\
   &  &  \ddots  &\ddots  &\\
   &  & & s_1 E_k  & B_k \end{array} \right),
   \end{gather}
where
\begin{gather*}
B_r = \left(\begin{array}{ccccc}
(s_2-R)\tilde E_0 & D_0  &  &  & \\
 & (s_2-R+1)\tilde E_1  & D_1 & &  \\
   &  &  \ddots  &\ddots  &\\
   &  & & s_2 \tilde E_R  & D_R \end{array} \right), \ R=0,1, \ldots, k, \\
 D_l = \left(\begin{array}{@{}c@{\,}c@{$\!\!\!\!$}c@{}c@{\ }c@{}}
(s_3-l) & j+k+1-s_1-s_2-s_3  &  &  & \\
 & s_3-l+1  & j+k-s_1-s_2-s_3 & &  \\
   &  &  \ddots  &\ddots  &\\
   &  & & s_3  & j+k+1-s_1-s_2-s_3-l \end{array} \right),
\end{gather*}
$l=0,1, \ldots, k$.

All rows of the matrix $A$ (A.25) are evidently linearly
independent.

Thus we had proved that the system of equations (A.4)--(A.6) is
non-degenerate in all cases listed in the formulae (A.7). Thus the
system (72) in the case $m=4$ is non-degenerate.

Considering only such systems of equations (A.4)--(A.6) that
correspond to $S_1=0$ we get a full set of non-linked subsystems
of the system (72) for $m=3$, and in the case $s_1=s_2=0$ we come
to full set of non-linked subsystems of the system (72) for $m=2$.
Consequently non-degeneracy of the system (72) for $m=2$ and $m=3$
follows from the adduced proof as a particular case.

Similarly (but with involvement of somewhat more cumbersome
calculations) it is possible to prove non-degeneracy of the
systems of linear algebraic equations (74) for coefficients of
Killing tensors of rank $j$ and order $s$.

\end{document}